\documentclass[reprint,amsmath,amssymb,aps,superscriptaddress]{revtex4-2}

\usepackage{graphicx}
\usepackage{dcolumn}
\usepackage{bm}
\usepackage{physics}
\usepackage{adjustbox}
\usepackage{xcolor}
\usepackage{ulem}

\begin{document}

\preprint{APS/123-QED}

\title{Topological charge and bulk-surface correspondence for quad-helicoid surface states in topological semimetals with two glide-time-reversal symmetries}

\author{Taiki Yukitake}
\affiliation{Department of Physics, Institute of Science Tokyo, 2-12-1 Ookayama,
Meguro-ku, Tokyo 152-8551, Japan}

\author{Shuichi Murakami}
\email{murakami@ap.t.u-tokyo.ac.jp}
\affiliation{Department of Applied Physics, University of Tokyo,
 7-3-1 Hongo, Bunkyo-ku, Tokyo 113-8656, Japan}
\affiliation{International Institute for Sustainability with Knotted Chiral Meta
Matter (WPI-SKCM$^2$), Hiroshima University, 1-3-1 Kagamiyama, Higashi-hiroshima, Hiroshima
739-8526, Japan}
\affiliation{Center for Emergent Matter Science, RIKEN, 2-1 Hirosawa, Wako, Saitama 351-0198, Japan}
\affiliation{Department of Physics, Institute of Science Tokyo, 2-12-1 Ookayama,
Meguro-ku, Tokyo 152-8551, Japan}

\date{\today}

\begin{abstract}
Quad-helicoid surface states (QHSSs) are unique surface states with two pairs of helicoid surface states in topological semimetals such as Dirac semimetals. So far, topologically protected QHSSs are shown to appear in spinless systems with two $\mathcal{GT}$ symmetries and $\mathcal{T}$ symmetry ($\mathcal{G}$: glide, $\mathcal{T}$: time-reversal). In this paper, we show that topologically protected QHSSs also appear in spinful/spinless systems with only two $\mathcal{GT}$ symmetries by defining new topological charges and establishing the bulk-surface correspondence. We first define a local $Z_2\times Z_2$ monopole charge for gapless nodes at $\mathcal{GT}$-invariant high-symmetry points and a global $Z_2$ charge reflecting the global topological feature of $\mathcal{GT}$-symmetric topological semimetals. Next, we show that the latter $Z_2$ classification corresponds to the presence or absence of QHSSs on the surface with two $\mathcal{GT}$ symmetries. In addition, we provide simplified formulas of the $Z_2$ charge under additional symmetries, and clarify some symmetry conditions where QHSSs are filling-enforced.
\end{abstract}

\maketitle

\section{\label{sec:Q1}Introduction}
Topological semimetals (SMs), including Weyl SMs~\cite{Murakami2007, Wan2011, Fang2012, Li2015, Xu2015, Lv2015, Tsirkin2017, HWang2020}, Dirac SMs~\cite{Young2012, Yang2014, Yang2015, Gao2016, Wang2012, Liu2014, Xu2014, Wang2013, Borisenko2014, Morimoto2014, Tang2016, Kargarian2016, Le2018, Kargarian2018, Wu2019, Lin2020, Wieder2020, Fang2021, Xia2022, Qian2023}, and nodal-line SMs~\cite{Burkov2011, Fang2015, Chen2016, Zhao2017, Li2018, Ahn2018, Bouhon2019, Sheng2019, Wang2020, Lee2020, Chen2021, Chen2022, Xue2023, Xiang2024, Ma2024, Wang2024, Yue2024}, have attracted many researchers due to their interesting physical properties. For topological SMs, it is essential to determine a topological charge that characterizes gapless nodes in the bulk Brillouin zone (BZ). It is because the charge protects the existence of gapless nodes under symmetry-preserving perturbations, and in many cases, leads to unique topological surface/hinge states~\cite{Wan2011, HWang2020, Wieder2020, Fang2021, Burkov2011, Zhao2017, Wang2020}. For example, in three-dimensional (3D) Weyl SMs, the monopole charge $C\in\mathbb{Z}$, defined as the Chern number on a sphere enclosing a Weyl point, protects the Weyl point and leads to helicoid surface states (HSSs), with the Fermi arcs at the Fermi energy~\cite{Wan2011, Li2015}. 

In contrast, in 3D Dirac SMs, the Dirac point is composed of two Weyl points with $C=1$ and $C=-1$, and the total charge is zero. Thus, in general, no topological surface states are expected~\cite{Kargarian2016, Le2018, Kargarian2018, Wu2019}. 
Meanwhile, in $\mathcal{GT}$-protected 3D Dirac SMs ($\mathcal{G}$: glide, $\mathcal{T}$: time-reversal), one can define $Z_2$ topological charges, which protect Dirac points and lead to unique surface states called multi-HSSs~\cite{Fang2016, Cheng2020, Cai2020, Su2022, Zhang2022, Hara2023, Zhang2023, Yukitake2025}. To be more specific, in Dirac SMs with one $\mathcal{GT}$ symmetry, a $Z_2$ monopole charge is defined for each Dirac point. Then the nontrivial value of the charge leads to double-helicoid surface states (DHSSs), which can be seen as the superposition of HSSs caused by $C=1$ and anti-HSSs caused by $C=-1$~\cite{Fang2016}. Moreover, in spinless Dirac SMs with $\mathcal{T}$ symmetry and two $\mathcal{GT}$ symmetries, another $Z_2$ charge is defined for a pair of two Dirac points. Then the nontrivial value of the charge leads to quad-helicoid surface states (QHSSs), which can be seen as the superposition of two HSSs and two anti-HSSs~\cite{Fang2016, Zhang2022}. Here, as opposed to the case of DHSSs, the previous studies assume that the system is spinless and has $\mathcal{T}$ symmetry to discuss QHSSs. Therefore, topologically protected QHSSs have been shown to appear only in spinless Dirac SMs with $\mathcal{T}$ and two $\mathcal{GT}$ symmetries.

In this paper, we show emergence of topologically protected QHSSs in spinful/spinless Dirac SMs without $\mathcal{T}$ but with two $\mathcal{GT}$ symmetries, by defining new topological charges and establishing the bulk-surface correspondence for QHSSs. Firstly, in Sec.~\ref{sec:Q2}, we briefly review previous studies on double/quad-HSSs in Dirac SMs with $\mathcal{GT}$ symmetries. Next, in Sec.~\ref{sec:Q3}, we define a local $Z_2\times Z_2$ monopole charge for each Dirac point at $\mathcal{GT}$-invariant high-symmetry points (HSPs) and a global $Z_2$ charge reflecting the global topological feature of $\mathcal{GT}$-symmetric Dirac SMs. Then, we discuss the relationship between the topological charges. In Sec.~\ref{sec:Q4}, we establish a bulk-surface correspondence, which claims that the $Z_2$ classification given by the global $Z_2$ charge in the bulk BZ corresponds to the presence or absence of QHSSs on the surface BZ with two $\mathcal{GT}$ symmetries. We also give some examples of gapless nodes leading to QHSSs based on the bulk-surface correspondence. In addition, in Sec.~\ref{sec:Q5}, we give some simplified formulas of the $Z_2$ charge under additional symmetries. Then we clarify some symmetry conditions where QHSSs are filling-enforced. In Sec.~\ref{sec:Q6}, we construct a tight-binding model with two $\mathcal{GT}$ symmetries to examine our results. We conclude our discussion in Sec.~\ref{sec:Q7}.

We note that, as we will discusse in the main text, our theory of QHSSs applies not only to $\mathcal{GT}$-symmetric Dirac SMs, but also to other topological SMs, such as Weyl SMs and nodal-line SMs, as long as the system has two $\mathcal{GT}$ symmetries, corresponding to the magnetic space group (MSG) \#32.138 ($Pb'a'2$).

\section{\label{sec:Q2}Review of previous studies}
 
\subsection{\label{sec:Q2-1}Double-helicoid surface states in Dirac systems with one $\mathcal{GT}$ symmetry}
Firstly, we briefly review studies on DHSSs in Dirac SMs with one $\mathcal{GT}$ symmetry. It is known that, in general, 3D Dirac SMs do not have any topological surface states around the projection of Dirac points on the surface BZ because the total Chern number for the Dirac point is zero~\cite{Kargarian2016, Le2018, Kargarian2018, Wu2019}. It is in contrast with 3D Weyl SMs, which have HSSs as a consequence of the nontrivial value of the monopole charge $C$ for Weyl points~\cite{Wan2011, Li2015}.

Meanwhile, recent studies show that, under $\mathcal{GT}$ symmetry, a new $Z_2$ monopole charge is defined for each Dirac point, and the charge leads to DHSSs~\cite{Fang2016, Yukitake2025}. To be more specific, we consider a Dirac SM with $\tilde{\Theta}_x=\mathcal{G}_x\mathcal{T}=\{M_x|\frac{1}{2}\frac{1}{2}0\}'$ on a primitive lattice, corresponding to MSG \#7.26 ($Pc'$). Moreover, we assume that some Dirac points exist on the high-symmetry lines (HSLs) $U:\{(u,\pi,\pi)|-\pi\le u\le\pi\}$ and $V:\{(u,\pi,0)|-\pi\le u\le\pi\}$, where $\tilde{\Theta}_x^2=-1$ holds. Then, for each Dirac point on $U$ or $V$, the $Z_2$ charge $\mathcal{Q}_x[S^2]$ defined as~\cite{Fang2016}
\begin{align}
    (-1)^{\mathcal{Q}_x[S^2]}=\frac{\mathrm{Pf}[\omega_x(K=0)]}{\sqrt{\mathrm{det}[\omega_x(K=0)]}}\frac{\mathrm{Pf}[\omega_x(K=\pi)]}{\sqrt{\mathrm{det}[\omega_x(K=\pi)]}} \label{eq:Q2-1}
\end{align}
becomes nontrivial, i.e., $\mathcal{Q}_x[S^2]=1\ (\mathrm{mod}\ 2)$.
Here, $S^2$ is a sphere enclosing the Dirac point, $K\in[-\pi,\pi]$ is the parameter for the cross section of $S^2$ by the plane $k_y=\pi$, the points $K=0,\pi$ are $\tilde{\Theta}_x$-invariant points on the cross section, $\omega_x (\bm{k})$ is the sewing matrix defined as $[\omega_x (\bm{k})]_{mn}=\mel{u_{m}(\tilde{\Theta}_x\bm{k})}{\tilde{\Theta}_x}{u_{n}(\bm{k})}\ (m,n=1,2,\dots N_{\mathrm{occ}})$, $\ket{u_{n}(\bm{k})}$ is the periodic part of the $n$-th occupied Bloch band, and $N_{\mathrm{occ}}$ is the total number of occupied bands. Furthermore, the nontrivial value of $\mathcal{Q}_x[S^2]$ leads to DHSSs around the projection of the Dirac point on the (001) surface BZ.

\begin{figure}[t]
\includegraphics[width=\columnwidth, pagebox=cropbox, clip]{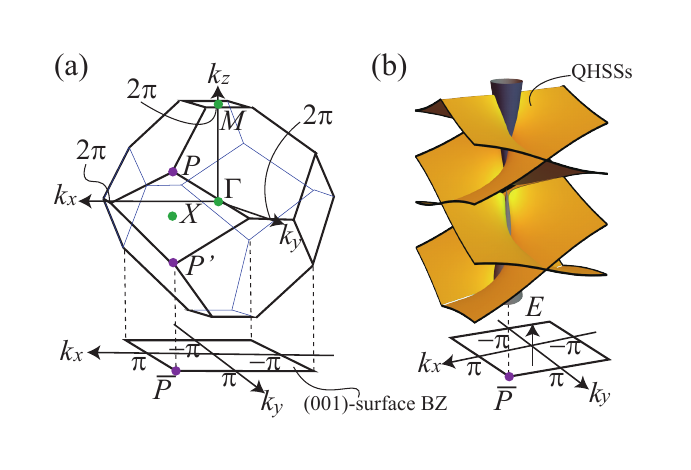}
\caption{\label{fig:Q1}\small{Quad-helicoid surface states. (a) Brillouin zone for MSG \#45.236 ($Iba21'$). We assume that Dirac points exist at high-symmetry points $P$ and $P'$.  (b) Quad-helicoid surface states. They can be seen as the superposition of two helicoid surface states and two anti-helicoid surface states. }}
\end{figure}

\subsection{\label{sec:Q2-2}Quad-helicoid surface states in Dirac systems with two $\mathcal{GT}$ symmetries}
Next, we briefly review studies on QHSSs in Dirac SMs with two $\mathcal{GT}$ symmetries. As we have seen above, in Dirac SMs with a single $\mathcal{GT}$ symmetry, Dirac points lead to DHSSs. Meanwhile, when two Dirac points in the bulk BZ are projected onto the same point in the surface BZ, we cannot expect the existence of DHSSs in general because both of the two Dirac points have $\mathcal{Q}_x[S^2]=1\ (\mathrm{mod}\ 2)$ and thus a pair of the two Dirac points have the trivial $Z_2$ charge in total~\cite{Fang2016}.

Meanwhile, previous studies show that such Dirac points can lead to other topological surface states called QHSSs in spinless Dirac SMs with $\mathcal{T}$ and two $\mathcal{GT}$ symmetries~\cite{Fang2016, Zhang2022}. To be more specific, we consider a spinless Dirac SM on a body-centered lattice with two glide symmetries $\mathcal{G}_x=\{M_x|\frac{1}{2}\frac{1}{2}0\},\ \mathcal{G}_y=\{M_y|\frac{1}{2}\frac{1}{2}0\}$, and time-reversal symmetry $\mathcal{T}$, corresponding to MSG \#45.236 ($Iba21'$). Moreover, we assume that only two Dirac points exist in the bulk BZ: one exists at the point $P: (\pi,\pi,\pi)$ and the other exists at the point $P':(\pi,\pi,-\pi)$ (see Fig.~\ref{fig:Q1}(a)). Then, the $Z_2$ charge $\mathcal{Q}_{xy}$ defined as~\cite{Zhang2022}
\begin{align}
(-1)^{\mathcal{Q}_{xy}}=\prod_{i\in\mathrm{occ}}\frac{\zeta_i(\Gamma)}{\zeta_i(X)}
\end{align}
becomes nontrivial. Here, $\zeta_i(\bm{k})$ is the $C_{2z}$ eigenvalue of the $i$-th occupied band at the $C_{2z}$-invariant $\bm{k}$, and $\Gamma :(0,0,0),\ X :(\pi,\pi,0)$ are HSPs in the bulk BZ for \#45.236. Furthermore, the nontrivial value of $\mathcal{Q}_{xy}$ leads to QHSSs around $\bar{P}:(\pi,\pi)$ in the (001) surface BZ, onto which the two Dirac points are projected (see Fig.~\ref{fig:Q1}(b)).

Here, we note that the discussion for QHSSs in Ref.~\cite{Zhang2022} applies only to spinless Dirac SMss with $\mathcal{T}$ and two $\mathcal{GT}$ symmetries. Therefore, it is not known whether QHSSs appear in spinful/spinless Dirac SMs with two $\mathcal{GT}$ symmetries but without $\mathcal{T}$ symmetry.

\section{\label{sec:Q3}Topological charge for quad-helicoid surface states}
While previous studies on QHSSs focused only on spinless Dirac SMs with $\mathcal{T}$ symmetry in addition to two $\mathcal{GT}$ symmetries, we find that we can define topological charges and establish bulk-surface correspondence for QHSSs in spinful/spinless Dirac SMs with only two $\mathcal{GT}$ symmetries.

In this section, we define a local $Z_2\times Z_2$ charge for each Dirac point at $\mathcal{GT}$-invariant HSPs and a global $Z_2$ charge reflecting the global topological feature of $\mathcal{GT}$-symmetric Dirac SMs.
To be more precise, we focus on a topological SMs, including Dirac SMs, with MSG \#32.138 ($Pb'a'2$) generated by translation operations $\{E |100\}, \{E |010\}, \{E |001\}$, and two $\mathcal{GT}$ symmetries $\tilde{\Theta}_x=\mathcal{G}_x\mathcal{T}=\{M_x|\frac{1}{2} \frac{1}{2} 0\}'$ and $\tilde{\Theta}_y=\mathcal{G}_y\mathcal{T}=\{M_y|\frac{1}{2} \frac{1}{2} 0\}'$. Moreover, we assume that gapless nodes such as Dirac points exist at HSPs $R:(\pi,\pi,\pi)$ and/or at $S:(\pi,\pi,0)$, where $\tilde{\Theta}_x^2=\tilde{\Theta}_y^2=-1$ holds (Fig.~\ref{fig:Q2}(a)). Then, we define topological charges for gapless nodes at the $\mathcal{GT}$-invariant HSPs.

\subsection{\label{sec:Q3-1}Local $Z_2\times Z_2$ charge $\mathcal{Q}_{xy}[S^2]$}
We first define a local $Z_2\times Z_2$ charge for Dirac points and other gapless nodes at $\mathcal{GT}$-invariant HSPs $R:(\pi,\pi,\pi)$ or $S:(\pi,\pi,0)$. 

\begin{figure}[t]
\includegraphics[width=\columnwidth, pagebox=cropbox, clip]{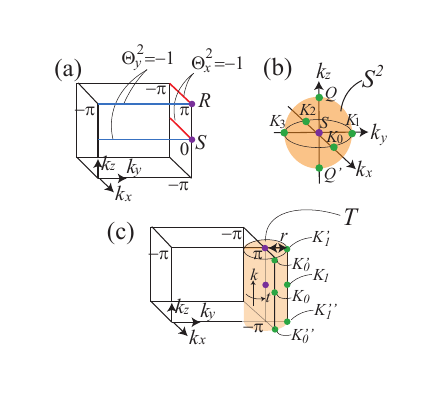}
\caption{\label{fig:Q2}\small{Topological charges for quad-helicoid surface states. (a) Brillouin zone for MSG \#32.138 ($Pb'a'2$). $R$ and $S$ are high-symmetry points with $\tilde{\Theta}_x^2=\tilde{\Theta}_y^2=-1$. (b) Sphere $S^2$ used to define $Z_2\times Z_2$ charge $\mathcal{Q}_{xy}[S^2]=(\mathcal{Q}_{xy}^{(0)}[S^2], \mathcal{Q}_{xy}^{(1)}[S^2])$ for $S$. $K_0, K_2$ are $\tilde{\Theta}_x$-invariant points on $S^2$, $K_1, K_3$ are $\tilde{\Theta}_y$-invariant points on $S^2$, and $Q, Q'$ are $C_{2z}$-invariant points on $S^2$. (c) Torus $T$ used to define $Z_2$ charge $\mathcal{Q}_{xy}[T]$. $K_0, K_0'$ are $\tilde{\Theta}_x$-invariant points on $T$, and $K_1, K_1'$ are $\tilde{\Theta}_y$-invariant points on $T$. }}
\end{figure}

To this end, we focus on a sphere $S^2$ centered at $R$ or $S$. We here assume that the sphere $S^2$ encloses targeted gapless nodes, such as a Dirac point at $R$ or $S$, while the system is gapped on $S^2$. Then, we define a $Z_2 \times Z_2$ monopole charge $\mathcal{Q}_{xy}[S^2]$ as
\begin{align}
\mathcal{Q}_{xy}[S^2]=(\mathcal{Q}_{xy}^{(0)}[S^2], \mathcal{Q}_{xy}^{(1)}[S^2])\in Z_2\times Z_2, 
\end{align}
with
\begin{align}
    &(-1)^{\mathcal{Q}_{xy}^{(0)}[S^2]}(-i)^{\frac{N_{\mathrm{occ}}}{2}f+n_-(Q)} \notag \\
    &= \frac{\sqrt{\mathrm{det}[\omega_x(Q)]}}{\sqrt{\mathrm{det}[\omega_y(Q)]}}\frac{\mathrm{Pf}[\omega_{x}(K_0)]}{\sqrt{\mathrm{det}[\omega_{x}(K_0)]}}\frac{\mathrm{Pf}[\omega_{y}(K_1)]}{\sqrt{\mathrm{det}[\omega_{y}(K_1)]}}, \label{eq:Q3-0}\\ 
    &(-1)^{\mathcal{Q}_{xy}^{(1)}[S^2]}(-i)^{\frac{N_{\mathrm{occ}}}{2}f+n_-(Q)} \notag \\
    &=\frac{\sqrt{\mathrm{det}[\omega_x(Q)]}}{\sqrt{\mathrm{det}[\omega_y(Q)]}}\frac{\mathrm{Pf}[\omega_{y}(K_1)]}{\sqrt{\mathrm{det}[\omega_{y}(K_1)]}}\frac{\mathrm{Pf}[\omega_{x}(K_2)]}{\sqrt{\mathrm{det}[\omega_{x}(K_2)]}}.
    \label{eq:Q3-1}
\end{align}
Here, the points $K_0,\ K_1$, and $Q$ are $\tilde{\Theta}_{x}$-, $\tilde{\Theta}_{y}$-, and $C_{2z}=\{E|\bar{1}00\}\tilde{\Theta}_y\tilde{\Theta}_x$-invariant points on $S^2$, respectively (see Fig.~\ref{fig:Q2}(b)), $f=0(1)$ when the system is spinless (spinful), $n_-(\bm{k})$ is the number of occupied bands with $C_{2z}=-i^f$ at the $C_{2z}$-invariant point $\bm{k}$, and the sewing matrix $\omega_{x_i}(\bm{k})$ is defined as $[\omega_{x_i}(\bm{k})]_{mn}=\mel{u_{m}(\tilde{\Theta}_{x_i}\bm{k})}{\tilde{\Theta}_{x_i}}{u_n(\bm{k})}$. 

Then, the charge $\mathcal{Q}_{xy}[S^2]$ gives a local $Z_2\times Z_2$ classification for gapless nodes at the HSPs $R$ or $S$. In fact, one can show that the charge is well-defined: gauge-independent, $\mathcal{Q}_{xy}^{(0)}[S^2]$ and $\mathcal{Q}_{xy}^{(1)}[S^2]$ are $Z_2$-valued ($\mathcal{Q}_{xy}^{(i)}[S^2]\in\{0,1\}$ for $i=0,1$), becomes trivial when there are no gapless nodes inside of $S^2$, is independent of the choice of HSPs on $S^2$, e.g., whether we choose $Q$ or $Q'$ in Eqs.~(\ref{eq:Q3-0}) and (\ref{eq:Q3-1}) does not affect the monopole charge (see Appendix~\ref{sec:QA-1}).  Moreover, one can find gapless nodes with $\mathcal{Q}_{xy}[S^2]=(0,1), (1,0)$, and $(1,1)$, respectively (see Sec.~\ref{sec:Q3-3}). Therefore, gapless nodes at $R$ or $S$ are classified into four types depending on the value of $\mathcal{Q}_{xy}[S^2]$, and gapless nodes with different values of $\mathcal{Q}_{xy}[S^2]$ cannot be transformed by perturbations mutually. In this sense, the charge $\mathcal{Q}_{xy}[S^2]$ gives the $Z_2\times Z_2$ classification for the gapless nodes.

We finally note that the value of $\mathcal{Q}_{xy}[S^2]$ depends on the choices of origin and symmetry axes of the system when $\mathcal{Q}[S^2]=1\ (\mathrm{mod}\ 2)$ (see Appendix~\ref{sec:QA-1}). Therefore, while we can use the charge to discuss the local classification of gapless nodes in $k$-space when we fix the choices, it is not directly related to physical responses in general.

\subsection{\label{sec:Q3-3}Examples of gapless nodes with nontrivial $\mathcal{Q}_{xy}[S^2]$}
We next give some examples of gapless nodes with the nontrivial value of the $Z_2\times Z_2$ charge $\mathcal{Q}_{xy}[S^2]$.

First, a Dirac point at $R$ or $S$ has $\mathcal{Q}_{xy}[S^2]=(1,0)$ or $(0,1)$. In fact, the new $Z_2\times Z_2$ charge $\mathcal{Q}_{xy}[S^2]$ is related to the previously defined $Z_2$ charges $\mathcal{Q}_{x_i}[S^2]\ (x_i=x,y)$ in Eq.~(\ref{eq:Q2-1}) as (see Appendix~\ref{sec:QA-3})
\begin{align}
    \mathcal{Q}[S^2]&\equiv\mathcal{Q}_{x}[S^2]=\mathcal{Q}_{y}[S^2] \notag \\
    &=\mathcal{Q}_{xy}^{(0)}[S^2]+\mathcal{Q}_{xy}^{(1)}[S^2] \notag \\
    &=n_-(Q)\ \ \ \ \ \ (\mathrm{mod}\ 2). \label{eq:Q3-7}
\end{align}
Then, since a Dirac point has the nontrivial value of $\mathcal{Q}[S^2]$, we obtain $\mathcal{Q}_{xy}[S^2]=(1,0)$ or $(0,1)$. Therefore, Dirac points at $\mathcal{GT}$-invariant HSPs are classified into two types based on the value of $\mathcal{Q}_{xy}[S^2]$. This fact is crucial for the emergence of QHSSs in $\mathcal{GT}$-symmetric Dirac SMs as we explain in Sec.~\ref{sec:Q4}. We finally note that, since a Weyl dipole~\cite{Zhang2022} (a pair of Weyl points related by $\tilde{\Theta}$) and a $Z_2$ nodal ring~\cite{Fang2015} also have $\mathcal{Q}[S^2]=1\ (\mathrm{mod}\ 2)$~\cite{Yukitake2025}, they also have $\mathcal{Q}_{xy}[S^2]=(1,0)$ or $(0,1)$.

\begin{figure}
\includegraphics[width=\columnwidth, pagebox=cropbox, clip]{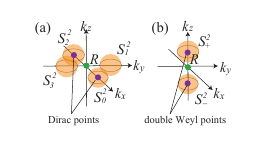}
\caption{\label{fig:Q5}\small{Example of gapless nodes with the nontrivial topological charge $\mathcal{Q}_{xy}[S^2]=(1,1)$. (a) Two Dirac points on the line $k_y=k_z=\pi$ related by $C_{2z}$ symmetry. (b) Two double Weyl points on the line $k_x=k_y=\pi$.  }}
\end{figure}

Second, a pair of Dirac points related by $C_{2z}$ symmetry has $\mathcal{Q}_{xy}[S^2]=(1,1)$ when $S^2$ encloses the pair (see Fig.~\ref{fig:Q5}(a)). In fact, since there are no gapless nodes on the line $k_x=k_y=\pi$ in such a case, we can transform the sphere $S^2$ into the union of four spheres $S^2_i\ (i=0,1,2,3)$, whose centers of are on $\tilde{\Theta}_{x_i}$-invariant lines, and we have (see Appendix~\ref{sec:QA-4})
\begin{align}
    \mathcal{Q}_{xy}^{(0)}[S^2]=\mathcal{Q}_{xy}^{(1)}[S^2]=\mathcal{Q}_{x}[S^2_0]+\mathcal{Q}_{y}[S^2_1]\ \ (\mathrm{mod}\ 2). \label{eq:Q3-8}
\end{align}
Then, since the pair of Dirac points related by $C_{2z}$ symmetry has $(\mathcal{Q}_{x}[S^2_0],\mathcal{Q}_{y}[S^2_1])=(1,0)$ or $(0,1)$, we have $\mathcal{Q}_{xy}^{(0)}[S^2]=\mathcal{Q}_{xy}^{(1)}[S^2]=1$. Therefore, the pair is topologically nontrivial, while it has the trivial value of the previously defined topological charge $\mathcal{Q}[S^2]$.

Third, a pair of double Weyl points~\cite{Fang2012} on the line $k_x=k_y=\pi$ has $\mathcal{Q}_{xy}[S^2]=(1,1)$, when $S^2$ encloses the pair (see Fig.~\ref{fig:Q5}(b)). This can be shown as follows. [Below, we focus on the HSP $R$.] When the system is gapped on the plane $k_z=\pi$, we can transform $S^2$ into the union of two spheres $S^2_{\pm}$ centered at $(\pi,\pi,\pi\pm\delta)$ with a constant $\delta>0$, and we have (see Appendix~\ref{sec:QA-5})
\begin{align}
    \mathcal{Q}_{xy}^{(0)}[S^2]&=\frac{C[S^2_+]-\Delta n_-}{2}\ (\mathrm{mod}\ 2), \label{eq:Q3-9}\\
    \mathcal{Q}_{xy}^{(1)}[S^2]&=\frac{C[S^2_+]+\Delta n_-}{2}\ (\mathrm{mod}\ 2), \label{eq:Q3-9-2}
\end{align}
where $C[S^2_+]$ is the Chern number on $S^2_+$, and $\Delta n_-=n_-(Q)-n_-(R)$ is the difference of $n_-$ between the two $C_{2z}$-invariant points on $S^2_+$. Then, since a double Weyl point enclosed by $S^2_+$ has $C[S^2_+]=\pm 2$, and it does not change the number of occupied bands with $C_{2z}=-i^f$, leading to $\Delta n_-=0$, we have $\mathcal{Q}_{xy}^{(0)}[S^2]=\mathcal{Q}_{xy}^{(1)}[S^2]=1$. We note that such a pair is realized in systems with spinful/spinless \#106.223 ($P4_2b'c'$) or spinless \#106.221 ($P4_2'b'c$) (see Sec.~\ref{sec:Q5-2}).

\subsection{\label{sec:Q3-2}Global $Z_2$ charge $\mathcal{Q}_{xy}[T]$}
So far, we introduced the local $Z_2\times Z_2$ classification for gapless nodes at $R$ or $S$, which is not directly related to any physical responses due to its origin/axes dependence. We next define a $Z_2$ charge that reflects the global topological feature of the $\mathcal{GT}$-symmetric topological SMs and does not have origin/axes dependence.

To this end, we next focus on a torus $T$ which simultaneously encloses the HSPs $R$ and $S$ (see Fig.~\ref{fig:Q2}(c)). We here assume that the torus $T$ encloses targeted gapless nodes, such as a pair of Dirac points at $R$ and $S$, while the system is gapped on $T$. Then we define a $Z_2$ charge  $\mathcal{Q}_{xy}[T]$ as
\begin{align}
    (-1)^{\mathcal{Q}_{xy}[T]}=\prod_{i=0,1}\frac{\mathrm{Pf}[\omega_{x_i}(K_i)]}{\sqrt{\mathrm{det}[\omega_{x_i}(K_i)]}}\frac{\mathrm{Pf}[\omega_{x_i}(K'_i)]}{\sqrt{\mathrm{det}[\omega_{x_i}(K'_i)]}}, \label{eq:Q4-1}
\end{align}
where $x_0=x$, $x_1=y$, $T$ is a torus defined as $T=\{\bm{k} | k_x=\pi+r\cos t,\ k_y=\pi+r\sin t,\ k_z=k  \ \ (-\pi\le k\le \pi\ ,-\pi\le t\le\pi)\}$ with constants $r(>0)$, $K_0, K_0'$ are $\tilde{\Theta}_{x}$-invariant points on $T$, and $K_1, K_1'$ are $\tilde{\Theta}_{x}$-invariant points on $T$ (see Fig.~\ref{fig:Q2}(c)). One can show that this $Z_2$ charge is also well-defined (see Appendix~\ref{sec:QB-1}). Moreover, one can see that the charge can become nontrivial. Therefore, the $Z_2$ charge $\mathcal{Q}_{xy}[T]$ gives the $Z_2$ classification of $\mathcal{GT}$-symmetric topological SMs.

The relationship between the $Z_2$ charge $\mathcal{Q}_{xy}[T]$ and the $Z_2\times Z_2$ charge $\mathcal{Q}_{xy}[S^2]$ is as follows. Let us consider two spheres $S^2(R)$ and $S^2(S)$ enclosing $R$ and $S$, respectively. Then, the union of two spheres can be smoothly transformed into the torus $T$ when there are no gapless nodes outside of the two spheres, and one can easily see
\begin{align}
    \mathcal{Q}_{xy}[T]&=\mathcal{Q}_{xy}^{(0)}[S^2(R)]+\mathcal{Q}_{xy}^{(0)}[S^2(S)] \notag \\
    &=\mathcal{Q}_{xy}^{(1)}[S^2(R)]+\mathcal{Q}_{xy}^{(1)}[S^2(S)]\ \  (\mathrm{mod}\ 2). \label{eq:Q3-6}
\end{align}
Therefore, the value of $\mathcal{Q}_{xy}[T]$ becomes nontrivial if and only if the value of $\mathcal{Q}_{xy}[S^2(R)]$ and the value of $\mathcal{Q}_{xy}[S^2(S)]$ are different: one is $(1,0)$ and the other is $(0,1)$, or one is $(1,1)$ and the other is $(0,0)$. 

Finally, the value of $\mathcal{Q}_{xy}[T]$ does not depend on the choices of origin and symmetry axes of the system. In fact, when $\mathcal{Q}[S^2(R)]=\mathcal{Q}[S^2(R)]=1\ (\mathrm{mod}\ 2)$, the values of $\mathcal{Q}^{(i)}_{xy}[S^2(R)]$ and $\mathcal{Q}^{(i)}_{xy}[S^2(S)]$ change simultaneously by one for some certain changes of origin/axes (see Appendix~\ref{sec:QA-1}). Meanwhile, when $\mathcal{Q}[S^2(R)]=\mathcal{Q}[S^2(R)]=0\ (\mathrm{mod}\ 2)$, the values of $\mathcal{Q}^{(i)}_{xy}[S^2(R)]$ and $\mathcal{Q}^{(i)}_{xy}[S^2(S)]$ do not charnge. Then, using Eq.~(\ref{eq:Q3-6}), we find that the value of $\mathcal{Q}_{xy}[T]$ is invariant. Therefore, it is expected that the charge $\mathcal{Q}_{xy}[T]$ has some physical meanings. In the next section, we show that this is indeed the case: the value of $\mathcal{Q}_{xy}[T]$ corresponds to the presence or absence of QHSSs.

\begin{figure*}[t]
\includegraphics[width=2\columnwidth, pagebox=cropbox, clip]{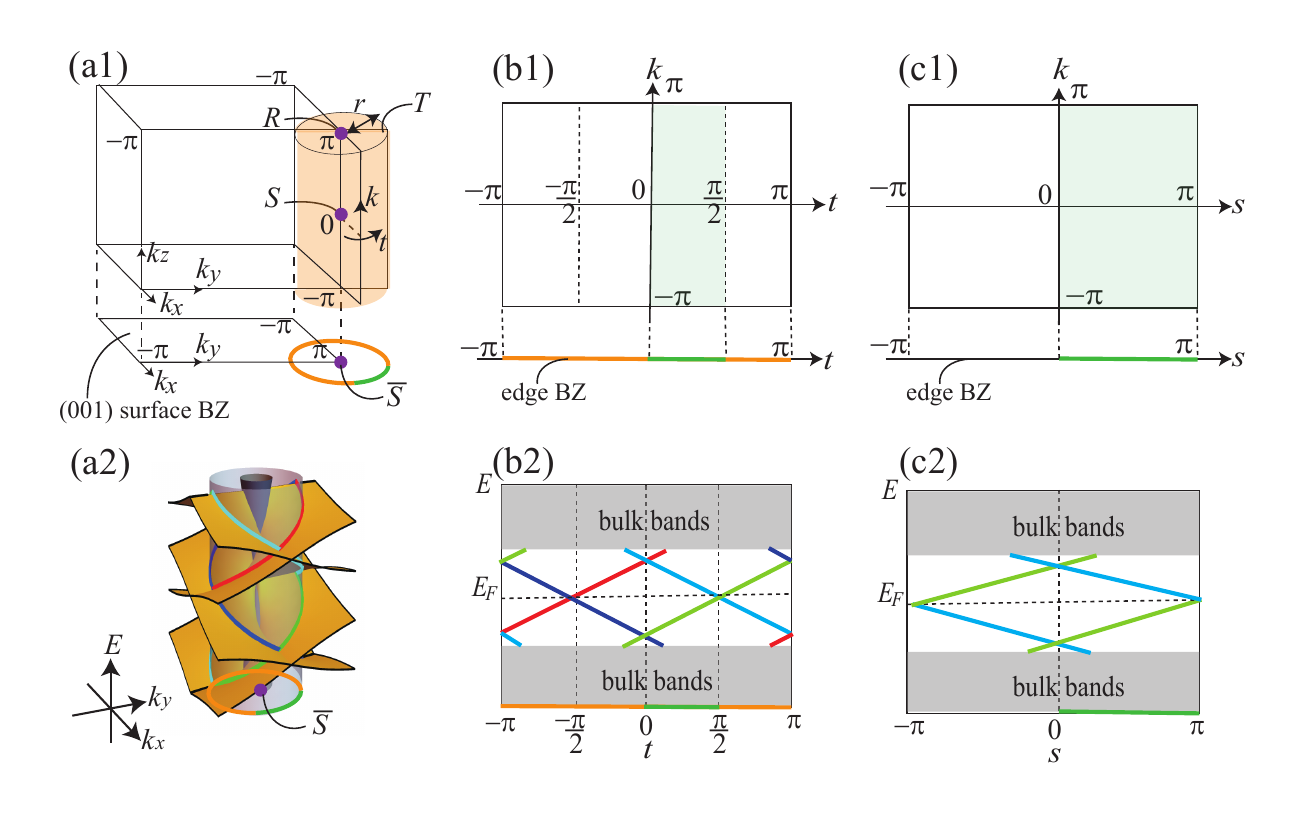}
\caption{\label{fig:Q3}\small{Bulk-surface correspondence for quad-helicoid surface states. (a1) Brillouin zone for MSG \#32.138 ($Pb'a'2$). The torus $T$ is projected onto the circle enclosing the point $(\pi,\pi)$ in the (001) surface Brillouin zone. (a2) Quad-helicoid surface states. They are obtained by considering the surface spectrum on the projection of the torus $T$ (yellow circle) with arbitrary values of the radius $r>0$. (b1) $(k,t)$-parameter space for the torus $T$. We can recognize this space as the 2D Brillouin zone for the 2D system $H(t,k)$ obtained by restricting the original 3D system to the torus. (b2) Edge spectrum of $H(t,k)$. When $\mathcal{Q}_{xy}[T]$ has a nontrivial value, then unique edge states appear, which can be considered as a section of QHSSs. (c1) 2D Brillouin zone for the 2D system $\tilde{H}(s,k)$. $\tilde{H}(s,k)$ is constructed from $H(t,k)$ such that $H(t,k)$ with $0\le t\le\pi/2$ and $\tilde{H}(s,k)$ with $0\le s\le\pi$ have the same bulk/edge spectra by identifying $s=2t$. (c2) Edge spectrum of $\tilde{H}(s,k)$. When $\mathcal{Q}_{xy}[T]$ has a nontrivial value, $\tilde{H}(s,k)$ can be seen as a $Z_2$ topological insulator, and thus helical edge states appear on the edge spectrum.} }
\end{figure*}

\section{\label{sec:Q4}Bulk-surface correspondence for quad-helicoid surface states}
 
As discussed in the previous section, in systems with two $\mathcal{GT}$ symmetries, the global topology of the bulk is characterized by the $Z_2$ charge $\mathcal{Q}_{xy}[T]$. In this section, we show that the value of $\mathcal{Q}_{xy}[T]$ corresponds to the presence or absence of QHSSs: we have $\mathcal{Q}_{xy}[T]=1\ (\mathrm{mod}\ 2)$ if and only if QHSSs appear around the projection of the torus $T$ in the (001) surface BZ in systems with two $\mathcal{GT}$ symmetries, corresponding to MSG \#32.138 ($Pb'a'2$). 

\subsection{\label{sec:Q4-1}Proof of the bulk-surface correspondence}
Firstly, we give a proof of the bulk-surface correspondence between $\mathcal{Q}_{xy}[T]$ and QHSSs. To prove the correspondence, we first regard the surface $T$ as a 2D BZ $(\mathbb{R}/2\pi\mathbb{Z})^2$ in the $(t,k)$ parameter space (see Sec.~\ref{sec:Q3-2}). Then by restricting the Bloch Hamiltonian $H(\bm{k})$ of the original 3D system to the torus $T$ (see Fig.~\ref{fig:Q3}(a1)), we obtain a 2D gapped system whose Bloch Hamiltonian is given by $H(t,k)$ (see Fig.~\ref{fig:Q3}(b1)). Next, by using $H(t,k)$, we construct another 2D gapped system $\tilde{H}(s,k)\ \ (\ (s,k)\in(\mathbb{R}/2\pi\mathbb{Z})^2\ )$ such that $H(t,k)\ (0\le t\le\frac{\pi}{2})$ and $\tilde{H}(s,k)\ (0\le s\le\pi)$ have the same bulk/edge spectra by identifying $s=2t$ (see Fig.~\ref{fig:Q3}(c1)), and $\tilde{H}(s,k)$ preserves an antiunitary symmetry $\tilde{\Theta}$ with $\tilde{\Theta}^2=-1$. The definition of $\tilde{H}(s,k)$ is given in Appendix~\ref{sec:QC}.  [The edge spectra mean the spectra of the edge states for the 1D systems whose Bloch Hamiltonians are obtained from $H(t,k)$ and $\tilde{H}(s,k)$, by identifying $k$ as the Bloch wavenumber.] 

As is similar to $Z_2$ topological insulators with time-reversal symmetry~\cite{FuKane2006, FuKane2007}, $\tilde{H}(s,k)$ can be classified by the $Z_2$ invariant $\nu$, whose nontrivial value leads to helical edge states on the edge spectrum. Here, one can easily see that the $Z_2$ invariant $\nu$ for $\tilde{H}(s,k)$ is equal to the $Z_2$ charge $\mathcal{Q}_{xy}[T]$ for the original 3D system $H(\bm{k})$. Therefore, $\tilde{H}(s,k)$ has helical edge states if and only if $\mathcal{Q}_{xy}[T]=1\ (\mathrm{mod}\ 2)$ (see Fig.~\ref{fig:Q3}(c2)). Then, since $\tilde{H}(s,k)$ with $0\le s\le\pi$ and $H(t,k)$ with $0\le t\le\frac{\pi}{2}$ have the same edge spectra by identifying $s=2t$, and the spectrum of $H(t,k)$ for the whole region is obtained by utilizing $\tilde{\Theta}_x$ and $\tilde{\Theta}_y$ symmetries, $H(t,k)$ has unique topological edge states which intersect the Fermi energy $E=E_F$ four times if and only if $\mathcal{Q}_{xy}[T]=1\ (\mathrm{mod}\ 2)$  (see Fig~\ref{fig:Q3}(b2)). Moreover, since $H(t,k)$ is defined by restricting the original 3D system $H(\bm{k})$ to the torus $T$, the edge states are identified with the surface states on the projection of $T$ in the (001) surface BZ. Here, as shown in Fig.~\ref{fig:Q3}(a2),  these surface states can be seen as a section of QHSSs. Therefore, by changing the radius $r>0$ of the torus $T$ within the range where the gap on $T$ does not close, we finally find that QHSSs appear around the projection of the torus $T$ in the (001) surface BZ if and only if $\mathcal{Q}_{xy}[T]=1\ (\mathrm{mod}\ 2)$. 

We finally note that, in general, since $\nu=1\ (\mathrm{mod}\ 2)$ if and only if edge states with $0\le t\le\pi$ intersect the Fermi energy odd times~\cite{Graf2013}, QHSSs are characterized as the surface states which intersect the Fermi energy $8n+4\ (n\in\mathbb{Z}_{\ge0})$ times on each circle centered at $(k_x,k_y)=(\pi,\pi)$ in the (001) surface BZ.

\subsection{\label{sec:Q4-2}Example of gapless nodes leading to quad-helicoid surface states}
We next give some examples of gapless nodes leading to QHSSs by using the newly defined topological charges and the bulk-surface correspondence.

First, in Dirac SMs on a primitive lattice, a pair of Dirac points at different $\mathcal{GT}$-invariant HSPs $R$ and $S$ leads to QHSSs when they have the opposite vale of $\mathcal{Q}_{xy}[S^2]$: one has $(1,0)$, and the other has $(0,1)$. In fact, in this case, we have $\mathcal{Q}_{xy}[T]=1\ (\mathrm{mod}\ 2)$ from Eq.~(\ref{eq:Q3-8}). Here, we can clarify whether the Dirac points at $R$ and $S$ have the opposite $\mathcal{Q}_{xy}[S^2]$ by adding a $\mathcal{GT}$-preserving perturbation. MSG \#32.138 only has 2D irreps at $R$ and $S$. Therefore, when we add a $\mathcal{GT}$-preserving perturbation, the Dirac point at $R$ splits into two Weyl points at $(\pi,\pi,\pi+\delta)$ and at $(\pi,\pi,\pi-\delta)$, and the Dirac point at $S$ splits into two Weyl points at $(\pi,\pi,\delta')$ and at $(\pi,\pi,-\delta')$, with constants $\delta,\delta'>0$. Then, we find from Eq.~(\ref{eq:Q3-9}) that, when only one of two quantities, $C(=\pm1)$ and $\Delta n_-(=\pm1)$, differs between the Weyl points at $(\pi,\pi,\pi+\delta)$ and at $(\pi,\pi,\delta')$, the two Dirac points have the opposite $\mathcal{Q}_{xy}[S^2]$, leading to the emergence of QHSSs.
We finally note that, in spinful Dirac SMs with MSG \#106.221 ($P4_2'b'c$), such a pair is symmetry-enforced as we will see in Sec.~\ref{sec:Q5-2}.

Second, similar to the first example, in Dirac SMs on a body-centered lattice, a pair of Dirac points at different $\mathcal{GT}$-invariant HSPs $W:(\pi,\pi,\pi)$ and $WA:(\pi,\pi,-\pi)$ leads to QHSSs when they have the opposite $Z_2\times Z_2$ charges. [The $Z_2\times Z_2$ charge $\mathcal{Q}_{xy}[S^2]$ can be defined also in body-centered systems (see Appendix~\ref{sec:QD-1}).]  Here, as we will see in Sec.~\ref{sec:Q5-3} and Appendix \ref{sec:QD-2}, in Dirac SMs with spinless \#44.233 ($I_cmm2$), spinless \#45.236 ($Iba21'$), spinful \#46.248 ($I_bma2$), spinful/spinless \#73.551 ($Ib'c'a$), spinful/spinless \#110.249 ($I4_1c'd'$), and spinless \#120.323 ($I\bar{4}'c'2$), additional symmetries force Dirac points at $W$ and $WA$ to have opposite $Z_2\times Z_2$ charges. Therefore, in Dirac SMs with such symmetries, a pair of Dirac points at $W$ and $WA$ always leads to QHSSs.

Third, in Weyl SMs on a primitive lattice, a pair of double Weyl points on the line $k_x=k_y=\pi$ leads to QHSSs. In fact, in this case, we have $\mathcal{Q}_{xy}[S^2(R)]=(1,1)$ and $\mathcal{Q}_{xy}[S^2(S)]=(0,0)$ when $S^2(R)$ encloses the pair (see Sec.~\ref{sec:Q3-3}). Then we have $\mathcal{Q}_{xy}[T]=1\ (\mathrm{mod}\ 2)$ by using Eq.~(\ref{eq:Q3-8}). As we will see in Sec.~\ref{sec:Q5-2}, in spinful/spinless Dirac SMs with MSG \#106.223 ($P4_2b'c'$) and spinless Dirac SMs with MSG \#106.221 ($P4_2'b'c$), such a pair is symmetry-enforced.

\section{\label{sec:Q5}Simplification of topological charge formula}

We next see that, with additional symmetries, we can simplify the formula of the $Z_2$ charge $\mathcal{Q}_{xy}[T]$, which is directly related to the emergence of QHSSs. This simplification would enable us to search for materials with QHSSs more easily, in the same way as the Fu-Kane formula for $Z_2$ topological insulators~\cite{FuKane2007}. Here, we focus on three types of symmetries: inversion symmetry (Sec.~\ref{sec:Q5-1}), fourfold symmetry (Sec.~\ref{sec:Q5-2}), and symmetry on a body-centered lattice (Sec.~\ref{sec:Q5-3}). 

\subsection{\label{sec:Q5-1}Inversion symmetry}
In this subsection, we see that $\mathcal{PT}$ and $\mathcal{P}$ symmetries can simplify the formula of $\mathcal{Q}_{xy}[T]$. We just give the results in the main text. We will provide the proofs in Appendices~\ref{sec:QB-3} and \ref{sec:QB-5}.

Firstly, we focus on spinful systems with $\mathcal{PT}$ and $\mathcal{GT}$ symmetries. In this case, we find that the formula of $\mathcal{Q}_{xy}[T]$ is expressed as the product of the eigenvalues of symmetry operators at HSPs in $k$-space, which is analogous to the original Fu-Kane formula for the $Z_2$ invariant $\nu$~\cite{FuKane2007}. To be more specific, systems with $\mathcal{PT}$ and $\mathcal{GT}$ are classified into three types: systems with \#50.283 ($Pb'a'n'$), \#54.345 ($Pc'c'a'$), or \#55.359 ($Pb'a'm'$), depending on the position of the inversion center. Then, for each type, the formula of $\mathcal{Q}_{xy}[T]$ is simplified as 
\begin{align}
       (-1)^{\mathcal{Q}_{xy}[T]}=\prod_{i=1}^{N_{\mathrm{occ}}/2}\frac{\zeta_{2i}[K_0]}{\zeta_{2i}[K_0']}\frac{\xi_{2i}[K_1]}{\xi_{2i}[K_1']}. \label{eq:Q5-1-1}  
\end{align}
Here, $\zeta_i(\bm{k})$ is the $C_{2x}=\{2_{100}|0\frac{1}{2}0\}$ (for \#50.283), $C_{2z}=\{2_{001}|\frac{1}{2}00\}$ (for \#54.345), or $\tilde{C}_{2x}=\{2_{100}|\frac{1}{2}\frac{1}{2}0\}$ (for \#55.359) eigenvalue of the $i$-th occupied band at $\bm{k}$, and $\xi_i(\bm{k})$ is the $C_{2y}=\{2_{010}|\frac{1}{2}00\}$ (for \#50.283), $\tilde{C}_{2x}=\{2_{100}|\frac{1}{2}0\frac{1}{2}\}$ (for \#54.345), or $\tilde{C}_{2y}=\{2_{010}|\frac{1}{2}\frac{1}{2}0\}$ (for \#55.359) eigenvalue of the $i$-th occupied band at $\bm{k}$. 

Next, we focus on spinful/spinless systems with $\mathcal{P}$ and $\mathcal{GT}$ symmetries. Systems with $\mathcal{P}$ and $\mathcal{GT}$ are also classified into three types: systems with \#50.281 ($Pb'a'n$), \#54.344 ($Pc'ca'$), or \#55.357 ($Pb'a'm$), depending on the position of the inversion center. Then, we find that, in systems with MSG \#55.357 ($Pb'a'm$), the $Z_2$ charge $\mathcal{Q}_{xy}[T]$ is always trivial: 
\begin{align}
    \mathcal{Q}_{xy}[T]=0\ \  (\mathrm{mod}\ 2). \label{eq:Q5-1-2}
\end{align}
Therefore, QHSSs are forbidden. Meanwhile, in systems with MSGs \#50.281 ($Pb'a'n$) or \#54.344 ($Pc'ca'$), as opposed to the case of \#55.357, $\mathcal{Q}_{xy}[T]$ can be nontrivial.

\subsection{\label{sec:Q5-2}Fourfold symmetry}
In this subsection, we see that fourfold rotation, screw, and rotoinversion symmetries simplify the formula of $\mathcal{Q}_{xy}[T]$. We just give the results in the main text. We will give the proofs in Appendix~\ref{sec:QB-6}.

Firstly, in systems with $C_{4z}=\{4^+_{001}|0 \}$, or with $C_{4z}\mathcal{T}=\{4^+_{001}|0 \}'$, corresponding to MSGs \#100.175 ($P4b'm'$) or  \#100.173 ($P4'b'm$), respectively, $\mathcal{Q}_{xy}[T]$ is always trivial:
\begin{align}
        \mathcal{Q}_{xy}[T]=0\ \ (\mathrm{mod}\ 2). \label{eq:Q5-2-1}
\end{align}
Therefore, QHSSs cannot appear. Here, we can naively understand the reason for the absence of QHSSs as follows. When QHSSs appear on the (001) surface BZ, the number of HSSs in QHSSs is equal to $8m+4$ (see Sec.~\ref{sec:Q4}). Meanwhile, when systems possess $C_{4z}$ or $C_{4z}\mathcal{T}$, the number of HSSs must be $8m$ ($4m$ HSSs and $4m$ anti-HSSs) due to the symmetries on the (001) surface, which contradicts the emergence of QHSSs. 

Secondly, in systems with $\tilde{C}_{4}=\{4^+_{001}|0 0 \frac{1}{2}\}$ or with $\tilde{C}_{4}\mathcal{T}=\{4^+_{001}|0 0 \frac{1}{2}\}'$, corresponding to MSGs \#106.223 ($P4_2b'c'$) or \#106.221 ($P4_2'b'c$), respectively, the formula of $\mathcal{Q}_{xy}[T]$ is simplified as
\begin{align}
    \mathcal{Q}_{xy}[T]=\frac{N_{\mathrm{occ}}}{2}\ \ (\mathrm{mod}\ 2). \label{eq:Q5-2-2}
\end{align}
Therefore, QHSSs are filling-enforced, i.e., QHSSs appear if and only if the filling $N_{\mathrm{occ}}$ is equal to $2\ \mathrm{mod}\ 4$. We note that, for spinful/spinless \#106.223 and spinless \#106.221, a pair of double Weyl points on the line $k_x=k_y=\pi$ leads to the QHSSs. This can be confirmed from Eq.~(\ref{eq:Q3-9}) and the irreducible co-representations at the HSPs $(\pi,\pi,0)$ and $(\pi,\pi,\pi)$. Meanwhile, for spinful \#106.221, Dirac points at the HSPs with opposite $Z_2\times Z_2$ charges $\mathcal{Q}_{xy}[S^2]=(0,1)$ and $(1,0)$ leads to the QHSSs.

Thirdly, in systems with $\bar{C_4}\mathcal{T}=\{\bar{4}^+_{001}|0\}'$, corresponding to MSG \#117.301 ($P\bar{4}'b'2$), $\mathcal{Q}_{xy}[T]$ is given by the Fu-Kane-like formula:
\begin{align}
    (-1)^{\mathcal{Q}_{xy}[T]}=\prod_{i=1}^{N_{\mathrm{occ}}}\frac{\alpha_i[S]}{\alpha_i[\Sigma]}. \label{eq:Q5-2-3} 
\end{align}
Here, $\alpha_i(\bm{k})$ is the $C_2=\{2_{110}|\frac{1}{2} \bar{\frac{1}{2}} 0\}$ eigenvalue of the $i$-th occupied band at $\bm{k}$, and $\Sigma, S$ are the $C_2$-invariant points on $T$, which correspond to $(k,t)=(\pi/4, 0), (\pi/4,\pi)$ under the $(k,t)$-parameterization for $T$ (see Sec.~\ref{sec:Q3-2}), respectively. Therefore, the band inversion on the lines $\{(u,u,0)|-\pi\le u\le\pi\}$ or $\{(u,u,\pi)|-\pi\le u\le\pi\}$ leads to QHSSs.

Finally, in systems with $\bar{C_4}=\{\bar{4}^+_{001}|0\}$, corresponding to MSG \#117.303 ($P\bar{4}b'2'$), $\mathcal{Q}_{xy}[T]$ is also given by the Fu-Kane-like formula:
\begin{align}
        (-1)^{\mathcal{Q}_{xy}[T]}=\prod_{i=1}^{N_{\mathrm{occ}}}\frac{\beta_i[Z]}{\beta_i[\Gamma]}, \label{eq:Q5-2-4} 
\end{align}
where $\beta_i(\bm{k})$ is the $\bar{C_4}$ eigenvalue of the $i$-th occupied band at $\bm{k}$, $\Gamma=(0,0,0)$, and $Z=(0,0,\pi)$. Here, as opposed to the other fourfold symmetries, we assumed that the system does not have gapless nodes on the plane $k_x=k_y$ except for the line $k_x=k_y=\pi$. Therefore, interestingly, QHSSs around the point $(k_x,k_y)=(\pi,\pi)$ in the (001) surface BZ are related to the $\bar{C_4}$-eigenvalues of occupied bands at HSPs away from the line $k_x=k_y=\pi$.

\subsection{\label{sec:Q5-3}Symmetry on a body-centered lattice}
In this subsection, we see that, in systems on a body-centered lattice, additional symmetries relating two $\mathcal{GT}$-invariant HSPs can simplify the formula of $\mathcal{Q}_{xy}[T]$. In the main text, we focus on Dirac SMs with two Dirac points at $W:(\pi,\pi,\pi)$ and $WA:(\pi,\pi,-\pi)$, similar to Sec.~\ref{sec:Q2-2}. In Appendix~\ref{sec:QD}, we discuss the case where Dirac points or some other gapless nodes exist at other points.

In Dirac SMs with two $\mathcal{GT}$ symmetries on a body-centered lattice, corresponding to MSG \#45.238 ($Ib'a'2$), since \#45.238 is a supergroup of \#32.138, one can also define the topological charges $\mathcal{Q}_{xy}[T]$ (see Appendix~\ref{sec:QD-1}). Then, we find that symmetries relating two $\mathcal{GT}$-invariant HSPs $W$ and $WA$ simplify the formula of $\mathcal{Q}_{xy}[T]$. Such symmetries are contained in MSGs \#44.233 ($I_cmm2$), \#45.236 ($Iba21'$), \#46.248 ($I_bma2$), \#72.543 ($Ib'a'm$), \#73.551 
($Ib'c'a$), \#108.237 ($I4c'm'$), \#110.249 ($I4_1c'd'$), \#120.323 ($I\bar{4}'c'2$), or the supergroups of these MSGs. They are classified into two groups $A$ and $B$ based on the simplification results as follows.

Group $A$: contains spinless \#44.233, spinless \#45.236, spinful \#46.248, spinful/spinless 73.551, spinful/spinless 110.249, and spinless \#120.323. Under these symmetries, the formula of $\mathcal{Q}_{xy}[T]$ is simplified as
\begin{align}
    \mathcal{Q}_{xy}[T]=n_-(X)\ \ (\mathrm{mod}\ 2), \label{eq:Q5-3-1}
\end{align}
where $n_-(X)$ is the number of occupied bands with $C_{2z}=-i^f$ at $X:(\pi,\pi,0)$. Here, the right-hand side of Eq.~(\ref{eq:Q5-3-1}) is equal to the $Z_2$ charge $\mathcal{Q}[S^2(W)]$, which becomes nontrivial for a Dirac point at $W$ (see Eq.~(\ref{eq:Q3-7})).
Therefore, interestingly, the pair of Dirac points at $W$ and $WA$ always leads to QHSSs. This is because symmetries in the group $A$ force the value of the $Z_2\times Z_2$ charge $\mathcal{Q}_{xy}[S^2]$ at $W$ and that at $WA$ to be different (see Appendix \ref{sec:QD-2}).
Moreover, since $n_-(X)=N_{\mathrm{occ}}/2$ for \#73.551 and \#110.249, QHSSs are filling-enforced in systems with the two MSGs:
\begin{align}
     \mathcal{Q}_{xy}[T]=\frac{N_{\mathrm{occ}}}{2}\ \ (\mathrm{mod}\ 2). \label{eq:Q5-3-2}
\end{align}

Group $B$: contains spinful \#44.233, spinful \#45.236, spinless \#46.248, spinful/spinless \#72.543, spinful/spinless \#108.237, and spinful \#120.323. Under these symmetries, the formula of $\mathcal{Q}_{xy}[T]$ is simplified as
\begin{align}
    \mathcal{Q}_{xy}[T]=0\ \ (\mathrm{mod}\ 2). \label{eq:Q5-3-3}
\end{align}
Therefore, in Dirac SMs with the MSGs, the pair of Dirac points at $W$ and $WA$ cannot lead to QHSSs. This is because symmetries in the group $B$ force the value of the $Z_2\times Z_2$ charge $\mathcal{Q}_{xy}[S^2]$ at $W$ and that at $WA$ to be the same (see Appendix \ref{sec:QD-2}). We note that, in systems with \#72.543 and \#108.237, $\mathcal{Q}_{xy}[T]$ is always trivial (see Eq.~(\ref{eq:Q5-1-2}) for \#72.543, and Eq.~(\ref{eq:Q5-2-1}) for \#108.237). Meanwhile, in systems with other MSGs, $\mathcal{Q}_{xy}[T]$ can become nontrivial when there are other gapless nodes.

Finally, we note that the results above are consistent with the results in previous studies. In fact, the $Z_2$ charge $\mathcal{Q}_{xy}$ defined in Ref.~\cite{Zhang2022} (see Sec~\ref{sec:Q2-2}) and our new charge $\mathcal{Q}_{xy}[T]$ are the same in spinless Dirac SMs with MSG \#45.236 ($Iba21'$): 
\begin{align}
    \mathcal{Q}_{xy}[T]=\mathcal{Q}_{xy}\ \ (\mathrm{mod}\ 2).
\end{align}
This can be show by using Eq.~(\ref{eq:Q5-3-1}) and the compatibility relations at $R: (\pi, 0, \pi)$, $S: (0,\pi,\pi)$, and $\Gamma:(0,0,0)$. Therefore, the bulk-surface correspondence for spinless systems with \#45.236 established in Ref.~\cite{Zhang2022} is included in the bulk-surface correspondence established in Sec.~\ref{sec:Q4}.

\begin{figure*}
\includegraphics[width=2\columnwidth, pagebox=cropbox, clip]{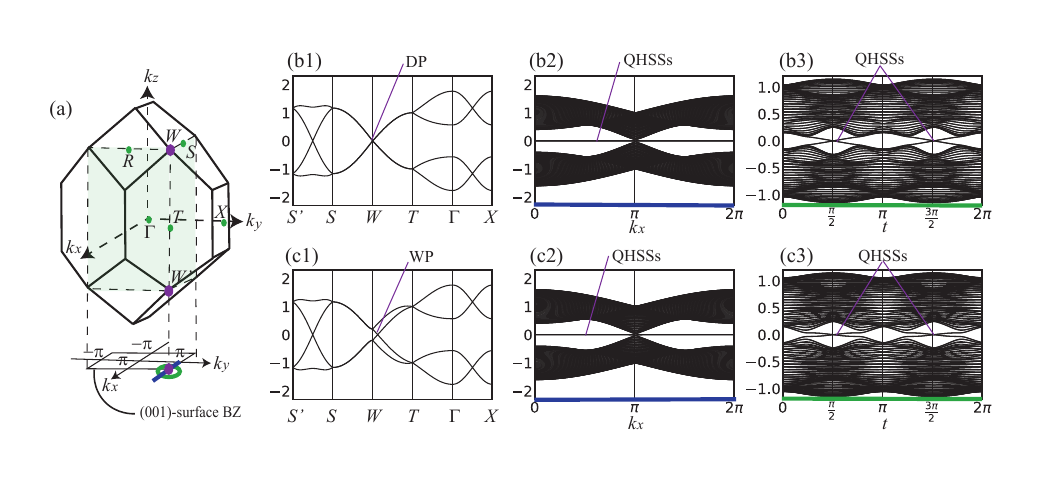}
\caption{\label{fig:Q4}\small{Band structures of $H_{\mathrm{TBM}}(\bm{k})$ and $H'_{\mathrm{TBM}}(\bm{k})$. (a) Brillouin zone for MSG \#73.551 ($Ib'c'a$). (b) Band structures of $H_{\mathrm{TBM}}(\bm{k})$. (b1) Bulk band structure. There is a Dirac point at $W:(\pi,\pi,\pi)$. Since the model has $\mathcal{P}$ symmetry, there is another Dirac point at $WA:(\pi,\pi,-\pi)$. (b2) (001) surface band structure on the blue line $(k_x,\pi)\ \  (0\le k_x\le 2\pi)$. QHSSs exist around the projection of Dirac points. (b3) (001) surface band structure on the green circle $(\pi+0.5\cos t,\pi+0.5\sin t)\ \  (0\le t\le 2\pi)$. The bands intersect the Fermi energy four times. This situation characterizes QHSSs. (c) Band structures of $H'_{\mathrm{TBM}}(\bm{k})=H_{\mathrm{TBM}}(\bm{k})+\Delta(\bm{k})$. (c1) Bulk band structure. Dirac points of $H_{\mathrm{TBM}}(\bm{k})$ split into pairs of two Weyl points. (c2) (001) surface band structure on the blue line $(k_x,\pi)\ \  (0\le k_x\le 2\pi)$. QHSSs exist around the projection of the pairs of Weyl points. (b3) (001) surface band structure on the green circle $(\pi+0.5\cos t,\pi+0.5\sin t)\ \  (0\le t\le 2\pi)$. Even after we add a perturbation, the bands intersect the Fermi energy four times.  } }
\end{figure*}

\section{\label{sec:Q6}Tight-binding model}
We next give a simple tight-binding model with two $\mathcal{GT}$ symmetries to examine our results. Here, we construct a spinless tight-binding model with MSG \#73.551 ($Ib'c'a$), which is generated by the translational symmetries for the body-centered lattice $\{E|100\}, \{E|010\},\{E|\frac{1}{2}\frac{1}{2}\frac{1}{2}\}$, two $\mathcal{GT}$ symmetries $\tilde{\Theta}_x=\{M_x|\frac{1}{2}\frac{1}{2}0\}',\ \tilde{\Theta}_y=\{M_y|\frac{1}{2}00\}'$ and the inversion symmetry $\mathcal{P}$. This MSG is in group A (see Sec.~\ref{sec:Q5-3}), and one can easily design the system to be a topologically nontrivial Dirac SM by using Eq.~(\ref{eq:Q5-3-1}). The Bloch Hamiltonian of the system is denoted by $H_{\mathrm{TBM}}(\bm{k})$, whose explicit form is given in Appendix~\ref{sec:QE}. Below, we assume half-filling, which means that two bands are occupied.

Let us observe the band structures of $H_{\mathrm{TBM}}(\bm{k})$. We first focus on the bulk band structure. As shown in Fig.~\ref{fig:Q4}(b1), $H_{\mathrm{TBM}}(\bm{k})$ has two Dirac points at the non-TRIM HSPs $W:(\pi,\pi,\pi)$ and $WA:(\pi,\pi,-\pi)$ related by $\mathcal{P}$. One can directly calculate the monopole charges $\mathcal{Q}[S^2(W)]$ and $\mathcal{Q}[S^2(WA)]$ for the Dirac points, and the $Z_2$ charge $\mathcal{Q}_{xy}[T]$ (see Appendix~\ref{sec:QE}). Then, we find $\mathcal{Q}[S^2(W)]=\mathcal{Q}[S^2(WA)]=\mathcal{Q}_{xy}[T]=1\ (\mathrm{mod}\ 2)$, which is consistent with Eqs.~(\ref{eq:Q5-3-1}) and~(\ref{eq:Q5-3-2}). Here, we note that other gapless nodes at (0,0,$\pm \pi)$ observed in Fig.~\ref{fig:Q4}(b1) are double Weyl points with $C=\pm 2$.

Next, we focus on the surface band structure. On the (001) surface BZ, as shown in Fig.~\ref{fig:Q4}(b2), gapless surface states appear around the point $(\pi,\pi)$, which is the projection of two Dirac points. Moreover, as shown in Fig.~\ref{fig:Q4}(b3), the surface states on the top surface intersect the Fermi energy $E_F=0$ four times. Therefore, these surface states are QHSSs, as expected from the bulk-surface correspondence in Sec.~\ref{sec:Q4}. 

Here, since all of the irreducible co-representations of \#73.551 at $W$ are two-dimensional, the Dirac point at $W$ is not symmetry-enforced. We next add a symmetry-preserving perturbation $\Delta(\bm{k})$ to the system (see Appendix~\ref{sec:QE}). Then, as shown in Fig.~\ref{fig:Q4}(c1), the two Dirac points at $W$ and $WA$ split into pairs of two Weyl points with opposite charge $C=\pm 1$. Even in this case, $\mathcal{Q}_{xy}[T]$ remains nontrivial and QHSSs still exist around the projection of these Weyl points as shown in Figs.~\ref{fig:Q4}(c2) and (c3).

\section{\label{sec:Q7}Conclusion}
In this paper, we showed that topologically protected quad-helicoid surface states appear in topological semimetals with two $\mathcal{GT}$ symmetries, corresponding to MSG \#32.138 ($Pb'a'2$). Firstly, we newly defined a local $Z_2\times Z_2$ monopole charge $\mathcal{Q}_{xy}[S^2]$, which gives the local classification of gapless nodes at the $\mathcal{GT}$-invariant high-symmetry points in the bulk Brillouin zone but depends on the choices of origin and symmetry axes of the system. We next defined a $Z_2$ charge $\mathcal{Q}_{xy}[T]$, which reflects the global topological feature of $\mathcal{GT}$-symmetric systems and does not have origin/axes dependence. Then we found that the $Z_2$ classification given by $\mathcal{Q}_{xy}[T]$ in the bulk Brillouin zone corresponds to the presence or absence of quad-helicoid surface states on the (001) surface Brillouin zone. In addition, we gave some simplified forms of the charge $\mathcal{Q}_{xy}[T]$ under additional symmetries. In particular, we found that, with MSGs \#106.223 ($P4_2b'c'$), \#106.221 ($P4_2'b'c$), \#73.551 ($Ib'c'a$), and \#110.249 ($I4_1c'd'$), the formula of $\mathcal{Q}_{xy}[T]$ is written using only the filling of the system, and thus QHSSs are filling-enforced on the (001) surface Brillouin zone.

In previous studies, quad-helicoid surface states have been observed experimentally only in spinless systems with $\mathcal{T}$ and two $\mathcal{GT}$ symmetries~\cite{Cheng2020, Cai2020}. Therefore, for future work, it is desired to find materials that break time-reversal symmetry but have quad-helicoid surface states. Moreover, it is also desired to uncover unique physical responses that appear as a consequence of the emergence of quad-helicoid surface states.

\begin{acknowledgments}
We acknowledge the support by Japan Society for the Promotion of Science (JSPS), KAKENHI Grant No. JP22K18687, No. JP22H00108, and No. JP24H02231.
\end{acknowledgments}

\appendix

\section{\label{sec:QA}Proof of some properties of $\mathcal{Q}_{xy}[S^2]$ }
In this section, we prove some properties of $\mathcal{Q}_{xy}^{(i)}[S^2]$ given in the main text. To this end, we rewrite the definition of $\mathcal{Q}_{xy}^{(i)}[S^2]$. Firstly, we fix a continuous gauge on the sphere $S^2$ centered at the HSP $R:(\pi,\pi,\pi)$ or $S:(\pi,\pi,0)$. One can take such a gauge because the Chern number on $S^2$ is equal to zero due to $\tilde{\Theta}_x$ symmetry. Next, we define $\tilde{d}_{x_i}: S^2\rightarrow S^1$ as a lift of $d_{x_i}: S^2\rightarrow S^1\ (\bm{k}\mapsto\mathrm{det}[\omega_{x_i}(\bm{k})])$ through the projection $p: S^1\rightarrow S^1\ (z\mapsto z^2)$, i.e., $\tilde{d}_{x_i}$ satisfies $\tilde{d}_{x_i}(\bm{k})^2=d_{x_i}(\bm{k})$. Here, $x_i=x$ for $i=0,2\ (\mathrm{mod}\ 4)$ and $x_i=y$ for $i=1,3\ (\mathrm{mod}\ 4)$. We can take such a lift because $p: S^2\rightarrow S^1$ is a covering map and $S^2$ is simply connected~\cite{Hatcher2001}. 
Then we can rewrite the definition of $\mathcal{Q}_{xy}^{(i)}[S^2]\ (i=0,1)$ as
\begin{align}
   &(-1)^{\mathcal{Q}_{xy}^{(i)}[S^2]} \notag \\
    &=i^{\frac{N}{2}f+n_-(Q)}\frac{\tilde{d}_x(Q)}{\tilde{d}_y(Q)}\frac{\mathrm{Pf}[\omega_{x_i}(K_i)]}{\tilde{d}_{x_i}(K_i)}\frac{\mathrm{Pf}[\omega_{x_{i+1}}(K_{i+1})]}{\tilde{d}_{x_{i+1}}(K_{i+1})}. \label{eq:QA-0}
\end{align}
Since there are only two lifts of $d_{x_i}$: $\tilde{d}_{x_i},\ -\tilde{d}_{x_i}$, the value of $\mathcal{Q}_{xy}^{(i)}[S^2]$ does not depend on the choice of lifts when we fix a continuous gauge. 

Here we note that Eq.~(\ref{eq:QA-0}) is meaningful also for $i=2,3$. Meanwhile, as we show below, we have $\mathcal{Q}_{xy}^{(2)}[S^2]=\mathcal{Q}_{xy}^{(0)}[S^2]\ (\mathrm{mod}\ 2)$ and $\mathcal{Q}_{xy}^{(3)}[S^2]=\mathcal{Q}_{xy}^{(1)}[S^2]\ (\mathrm{mod}\ 2)$. Therefore, only two independent quantities are obtained from Eq.~(\ref{eq:QA-0}), and thus we can define the $Z_2\times Z_2$ charge as in the main text.

\subsection{\label{sec:QA-1}Well-definedness of $\mathcal{Q}_{xy}^{(i)}[S^2]$}
In this subsection, we show that $\mathcal{Q}_{xy}^{(i)}[S^2]$ is well-defined, as claimed in Sec.~\ref{sec:Q3-1}.

Firstly, $\mathcal{Q}_{xy}^{(i)}[S^2]\ (i=0,1,2,3)$ is gauge-independent. To show this, we consider $U(N_{\mathrm{occ}})$ gauge transformation $U(\bm{k})\ (\ket{u_{m\bm{k}}}\mapsto \ket{u_{n\bm{k}}}'=U(\bm{k})_{mn}\ket{u_{m\bm{k}}}\ (m,n=1,2,\dots, N_{\mathrm{occ}}))$ on the sphere $S^2$, and define $\tilde{d}_{U^*}: S^2\rightarrow S^1$ as a lift of $d_{U^*}: S^2\rightarrow S^1\ (\bm{k}\mapsto \det[U(\bm{k})^*])$. Then the equation $\omega_{x_i}(\bm{k})^U=U(\tilde{\Theta}_{x_i}\bm{k})^{\dagger}\omega_{x_i}(\bm{k})U(\bm{k})^{*}$ leads to $\tilde{d}_{x_i}^U(\bm{k})=\tilde{d}_{U^*}(\tilde{\Theta}_{x_i}\bm{k})\tilde{d}_{U^*}(\bm{k})\tilde{d}_{x_i}(\bm{k})$, and $\mathrm{Pf}[\omega_{x_i}(K_i)]^U=\det[U(K_i)^*]\mathrm{Pf}[\omega _{x_i}(K_i)]$. [Here, $()^U$ means that the quantity is calculated by using the gauge transformed by $U(\bm{k})$.] Using these relations, one can easily see $(\mathcal{Q}_{xy}^{(i)}[S^2])^U=\mathcal{Q}_{xy}^{(i)}[S^2]\ \ (\mathrm{mod}\ 2)$. 

Secondly, $\mathcal{Q}_{xy}^{(i)}[S^2]$ is $Z_2$-valued. To show this, we use the relation $\tilde{\Theta}_x=\tilde{\Theta}_yC_{2z}$, which leads to
$\omega_x(\bm{k})=\omega_y(C_{2z}\bm{k})\omega_{C_{2z}}(\bm{k})^*$. Then, one can see
\begin{align}
    \qty(i^{\frac{N_{\mathrm{occ}}}{2}f+n_-(Q)}\frac{\tilde{d}_x(Q)}{\tilde{d}_y(Q)})^2=\mathrm{det}[\omega_{C_{2z}}(Q)]\frac{\mathrm{det}[\omega_x(Q)]}{\mathrm{det}[\omega_y(Q)]}=1. \label{eq:QA-1}
\end{align}
Meanwhile, since $\mathrm{Pf}^2=\mathrm{det}$ and $\tilde{d}_{x_i}^2=d_{x_i}$, we have
\begin{align}
    \qty(\frac{\mathrm{Pf}[\omega_{x_i}(K_i)]}{\tilde{d}_{x_i}(K_i)})^2 &=1. \label{eq:QA-2}
\end{align}
Combining Eqs.~(\ref{eq:QA-1}) and (\ref{eq:QA-2}), we find $(-1)^{2\mathcal{Q}_{xy}^{(i)}[S^2]}=1$ and thus $\mathcal{Q}_{xy}^{(i)}[S^2]\in \{0,1\}$.

Thirdly, $\mathcal{Q}_{xy}^{(i)}[S^2]=0$ when there are no gapless nodes inside of the sphere $S^2$. Below, we focus on the HSP $R$. Almost the same discussion holds for the other HSP $S$. First, since $\mathcal{Q}_{xy}^{(i)}[S^2]$ is a well-defined $Z_2$-valued quantity, we can shrink $S^2$ without changing the value of $\mathcal{Q}_{xy}^{(i)}[S^2]$ unless $S^2$ passes through gapless nodes. Therefore, we can set $r=0$, which leads to
\begin{align}
    (-1)^{\mathcal{Q}_{xy}^{(i)}[S^2]}=i^{\frac{N_{\mathrm{occ}}}{2}f+n_-(R)}\frac{\mathrm{Pf}[\omega_x(R)]}{\mathrm{Pf}[\omega_y(R)]}. \label{eq:QA-3}
\end{align}
Meanwhile, using the relation $\tilde{\Theta}_x=\tilde{\Theta}_yC_{2z}$, one can see that there are two irreducible co-representations at $R$, both of them are two-dimensional, and they are distinguished by the eigenvalue of $C_{2z}$. Therefore, decomposing the occupied bands at $R$ into the direct sum of irreducible co-representations, we have
\begin{align}
    \mathrm{Pf}[\omega_y(R)]=(i^f)^{n_+(R)/2}(-i^f)^{n_-(R)/2}\mathrm{Pf}[\omega_x(R)]. \label{eq:QA-4}
\end{align}
Then, combining Eqs.~(\ref{eq:QA-3}) and (\ref{eq:QA-4}), we finally find $\mathcal{Q}_{xy}^{(i)}[S^2]=0$. 

Fourthly, the value of $\mathcal{Q}_{xy}^{(i)}[S^2]$ does not depend on the choice of $Q$. The point $Q$ is defined as a $C_{2z}$-invariant point on $S^2$, and there are two such points. We refer to one of the points as $Q$ and the other as $Q'$. Then, using $\tilde{\Theta}_xC_{2z}=(-1)^f\{E|110\}C_{2z}\tilde{\Theta}_x$, we have $n_-(Q')=n_-(Q)$. Meanwhile, using $\tilde{\Theta}_x^2=\{E|010\}$ and $\tilde{\Theta}_y^2=\{E|100\}$, we have $\tilde{d}_{x_i}(Q')=\tilde{d}_{x_i}(Q)$. In fact, the relations lead to $\tilde{d}_{x_i}(\tilde{\Theta}_{x_i}\bm{k})=s\tilde{d}_{x_i}(\bm{k})\ (s=\pm 1)$ on the plane $k_{x_{i+1}}=\pi$, and we find $s=1$ because this equation holds with $s=1$ when $\bm{k}=K_i$. Therefore, we can use $Q'$ instead of $Q$ in Eq.~(\ref{eq:QA-0}).

Finally, we can use the pair $(K_2, K_3)$ instead of $(K_0, K_1)$ in Eq.~(\ref{eq:Q3-0}), and $(K_3, K_0)$ instead of $(K_1, K_2)$ in Eq.~(\ref{eq:Q3-1}) to define $\mathcal{Q}_{xy}^{(i)}[S^2]$. In fact, as we discuss in Appendix~\ref{sec:QA-3}, we have
\begin{align}
    \mathcal{Q}_{xy}^{(i+1)}[S^2]=\mathcal{Q}_{xy}^{(i)}[S^2]+n_-(Q)\ \ (\mathrm{mod}\ 2), \label{eq:QA-5}
\end{align}
for $i=0,1,2,3\ (\mathrm{mod}\ 4)$. Then, using this equation, we obtain $\mathcal{Q}_{xy}^{(i+2)}[S^2]=\mathcal{Q}_{xy}^{(i)}[S^2]\ \ (\mathrm{mod}\ 2)$.

Before moving on to the next subsection, we note that the value of $\mathcal{Q}_{xy}[S^2]$ depends on choices of origin and symmetry axes of the system when $\mathcal{Q}[S^2]=1\ (\mathrm{mod}\ 2)$. Here, we show the dependence on the origin. [We assume that one of the rotation axes passes through the new origin.] When the origin of the system is shifted by $\bm{a}$ with $2\bm{a}\in\mathbb{Z}^3$, $\tilde{\Theta}_{x_i}$ is changed as $\tilde{\Theta}_{x_i}'=\{E|\bm{a}\}\tilde{\Theta}_{x_i}\{E|\bm{a}\}^{-1}=e^{i\theta_i(\bm{k})}\tilde{\Theta}_{x_i}$ with $\theta_{i}(\bm{k})=-\bm{\tilde{\Theta}_{x_i}k}\cdot\bm{a}-\bm{k}\cdot\bm{a}$. Then, when we use $\tilde{\Theta}_{x_i}'$ instead of $\tilde{\Theta}_{x_i}$, we have
\begin{align}
    \tilde{d}_{x_i}(\bm{k})'&=e^{i\frac{N_{\mathrm{occ}}}{2}\theta_{i}(\bm{k})}\tilde{d}_{x_i}(\bm{k}), \\
    \mathrm{Pf}[\omega_x(K_i)]'&=e^{i\frac{N_{\mathrm{occ}}}{2}\theta_{i}(\bm{k})}\mathrm{Pf}[\omega_x(K_i)],
\end{align}
where $()'$ means that the quantity is calculated with $\tilde{\Theta}_{x_i}'$. Moreover, one can see $n_-(Q)'=n_-(Q)$ when $s=\theta_0(Q)+\theta_1(Q)=0\ (\mathrm{mod}\ 2\pi)$, and $n_-(Q)'=N_{\mathrm{occ}}-n_-(Q)$ when $s=\pi\ (\mathrm{mod}\ 2\pi)$. Then, combining these relations, we finally obtain
\begin{align}
     \mathcal{Q}_{xy}^{(i)}[S^2]'=\mathcal{Q}_{xy}^{(i)}[S^2]+\frac{s}{\pi}\mathcal{Q}[S^2]\ \ \  (\mathrm{mod}\ 2).
\end{align}
Therefore, the value of $\mathcal{Q}_{xy}^{(i)}[S^2]$ changes by one if $s=\theta_0(Q)+\theta_1(Q)=\pi\ (\mathrm{mod}\ 2\pi)$ and $\mathcal{Q}[S^2]=1\ (\mathrm{mod}\ 2)$.

\subsection{\label{sec:QA-3}Proof of Eq.~(\ref{eq:Q3-7})} 
In this subsection, we give a proof of Eq.~(\ref{eq:Q3-7}) in Sec.~\ref{sec:Q3-3}. Firstly, the relation $\tilde{\Theta}_xC_{2z}=(-1)^f\{E|110\}C_{2z}\tilde{\Theta}_x$ leads to
\begin{align}
    \omega_{x}(C_{2z}\bm{k})\omega_{C_{2z}}(\bm{k})^*=(-1)^fe^{ik_x-ik_y}\omega_{C_{2z}}(\tilde{\Theta}_x\bm{k})\omega_x(\bm{k}). \label{eq:QA-6}
\end{align}
Then, from Eq.~(\ref{eq:QA-6}), one can find
\begin{align}
    \tilde{d}_x(C_{2z}\bm{k})&=se^{i\frac{N_{\mathrm{occ}}}{2}(k_x-k_y)}\tilde{d}_{C_{2z}}(\tilde{\Theta}_x\bm{k})\tilde{d}_{C_{2z}}(\bm{k})\tilde{d}_x(\bm{k}), \label{eq:QA-7}\\
    \mathrm{Pf}[\omega_x(K_2)]&=(-1)^{\frac{N_{\mathrm{occ}}}{2}f}e^{i\frac{N_{\mathrm{occ}}}{2}r}\mathrm{det}[\omega_{C_{2z}}(K_0)]\mathrm{Pf}[\omega_x(K_0)], \label{eq:QA-8}
\end{align}
with $s=\pm 1$. Here, $\tilde{d}_{C_{2z}}:S^2\rightarrow S^1$ is a lift of $d_{C_{2z}}: S^2\rightarrow S^1\ (\bm{k}\mapsto\mathrm{det}[\omega_{C_{2z}}(\bm{k})])$ through the projection $p$. Then, the substitution of Eqs.~(\ref{eq:QA-7}) and (\ref{eq:QA-8}) into Eq.~(\ref{eq:QA-0}) leads to
\begin{align}
    (-1)^{\mathcal{Q}_{xy}^{(1)}[S^2]}=\tilde{d}_{C_{2z}}(Q)\tilde{d}_{C_{2z}}(Q')(-1)^{\mathcal{Q}_{xy}^{(0)}[S^2]+\frac{N_{\mathrm{occ}}}{2}f}.
\end{align}
Moreover, as we show below, we have $\tilde{d}_{C_{2z}}(Q')=\tilde{d}_{C_{2z}}(Q)$ and thus $\tilde{d}_{C_{2z}}(Q)\tilde{d}_{C_{2z}}(Q')=\det[B_{C_{2z}}(Q)]=(-1)^{N_{\mathrm{occ}}f/2+n_-(Q)}$. Therefore, we finally obtain Eq.~(\ref{eq:Q3-7}).

One can show $\tilde{d}_{C_{2z}}(Q')=\tilde{d}_{C_{2z}}(Q)$ as follows. 
Firstly, since $d_{C_{2z}}(Q')=d_{C_{2z}}(Q)$, we can consider the winding number of $d_{C_{2z}}\circ l$ denoted by $N_{{C_{2z}}}[l]$, where $l$ is a path from $Q$ to $Q'$ on $S^2$ parameterized by $t\in[-\pi,\pi]$. Then the equation $d_{C_{2z}}(C_{2z}\bm{k})=d_{C_{2z}}(\bm{k})^*$ leads to $N_{{C_{2z}}}[l']=-N_{{C_{2z}}}[l]$ with $l'(t)=C_{2z}l(t)$. Meanwhile, we have $N_{{C_{2z}}}[l]=N_{{C_{2z}}}[l']$ because $d_{C_{2z}}\circ l$ is homotopic to $d_{C_{2z}}\circ l'$. Therefore, we find $N_{{C_{2z}}}[l]=0$. Then, since $\tilde{d}_{C_{2z}}(Q)$ and $\tilde{d}_{C_{2z}}(Q')$ are related as $\tilde{d}_{C_{2z}}(Q')=(-1)^{N_{{C_{2z}}}[l]}\tilde{d}_{C_{2z}}(Q)$~\cite{Hatcher2001}, we finally obtain $\tilde{d}_{C_{2z}}(Q)=\tilde{d}_{C_{2z}}(Q')$.

We finally note that almost the same discussion leads to Eq.~(\ref{eq:QA-5}).

\begin{figure}
\includegraphics[width=\columnwidth, pagebox=cropbox, clip]{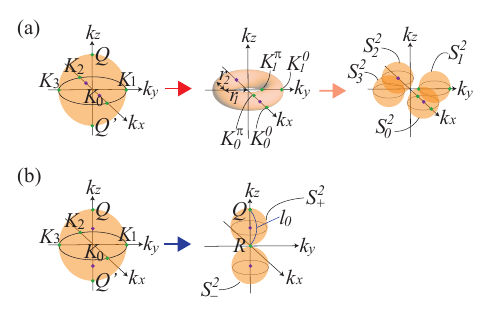}
\caption{\label{fig:QA}\small{Transformation of the sphere $S^2$. (a) When there are no gapless nodes on the line $k_x=k_y=\pi$, $S^2$ is transformed into a torus $\{\bm{k}| k_x=\pi+r_1\cos t+r_2\cos s\cos t,\ k_y=\pi+r_1\sin t+r_2\cos s\sin t,\ k_z=\pi+r_2\sin s\ (-\pi\le s,t\le\pi)\}$ (red arrow). Moreover, when gapless nodes exist only on the lines $k_y=k_z=\pi$ and $k_x=k_z=\pi$, we can transform the torus into the union of four spheres $S^2_i\ (i=0,1,2,3)$ (orange arrow). (b) When there are no gapless nodes on the plane $k_z=\pi$, $S^2$ is transformed into the union of two spheres $S^2_{\pm}$ (blue arrow).}}
\end{figure}

\subsection{\label{sec:QA-4}Proof of Eq.~(\ref{eq:Q3-8})}
In this subsection, we give a proof of Eq.~(\ref{eq:Q3-8}) in Sec.~\ref{sec:Q3-3}. Here, we focus on the HSP $R$. Almost the same discussion holds for the other HSP $S$. Firstly, when there are no gapless nodes on the line $k_x=k_y=\pi$, one can transform $S^2$ into a torus $\{\bm{k}| k_x=\pi+r_1\cos t+r_2\cos s\cos t,\ k_y=\pi+r_1\sin t+r_2\cos s\sin t,\ k_z=\pi+r_2\sin s\ (-\pi\le s,t\le\pi)\}$ with constants $0<r_2\le r_1\le r/2$ (see Fig.~\ref{fig:QA}(a)), without changing the value of $\mathcal{Q}_{xy}[S^2]$. Here, when $r_1=r_2=r/2$, we can relate $n_-(Q)=n_-(R)$ with $\mathrm{Pf}[\omega_{x_i}(R)]$ by using Eq.~(\ref{eq:QA-4}). Therefore, for the torus, we have
\begin{align}
    (-1)^{\mathcal{Q}^{(i)}_{xy}[S^2]}=\prod_{j=0,1}\frac{\mathrm{Pf}[\omega_{x_j}(K^0_j)]}{\tilde{d}_{x_j}(K^0_j)}\frac{\mathrm{Pf}[\omega_{x_j}(K^{\pi}_j)]}{\tilde{d}_{x_j}(K^{\pi}_j)}, \label{eq:QA-11}
\end{align}
where $K_0^0,\ K_0^{\pi},\ K_1^0,$ and $K_1^{\pi}$ are HSPs on the torus corresponding to $(s,t)=(0,0),(\pi,0),(0,\pi/2)$, and $(\pi,\pi/2)$ under the parametrization, respectively. Next, when gapless nodes exist only on the lines $k_y=k_z=\pi$ and $k_x=k_z=\pi$, we can transform the torus into the union of four spheres $S^2_i\ (i=0,1,2,3)$ centered at $(\pi+r_1,\pi,\pi)$, $(\pi,\pi+r_1,\pi)$, $(\pi-r_1,\pi,\pi)$, and $(\pi,\pi-r_1,\pi)$, respectively (see Fig.~\ref{fig:QA}(a)). Then, since the value of the right-hand side of Eq.~(\ref{eq:QA-11}) remains unchanged under the transformation, we finally obtain Eq.~(\ref{eq:Q3-8}) from Eq.~(\ref{eq:Q2-1}).

\subsection{\label{sec:QA-5}Proof of Eqs.~(\ref{eq:Q3-9}) and (\ref{eq:Q3-9-2})}
In this subsection, we give a proof of Eqs.~(\ref{eq:Q3-9}) and (\ref{eq:Q3-9-2}). Below, we focus on the HSP $R:(\pi,\pi,\pi)$. Almost the same discussion holds for the other HSP $S:(\pi,\pi,0)$. 

Firstly, we give a proof of Eq.~(\ref{eq:Q3-9}). To this end, we take a smooth gauge on the half part of $S^2(R)$ with $k_y\ge\pi$. [One can easily see that the gauge is enough to calculate the value of $\mathcal{Q}^{(i)}_{xy}[S^2]$.] Next, when there are no gapless nodes on the plane $k_z=\pi$, we can transform $S^2(R)$ into the union of two spheres $S^2_{\pm}=\{\bm{k}=(\pi+\frac{r}{2}\sin\theta\cos\phi, \pi+\frac{r}{2}\sin\theta\sin\phi, \pi+\frac{r}{2}(\pm1+\cos\theta))|\ 0\le\theta\le\pi,\ 0\le\phi\le 2\pi\}$ (see Fig.~\ref{fig:QA}(b)). Then, by using Eq.~(\ref{eq:QA-4}), we have
\begin{align}
    (-1)^{\mathcal{Q}^{(i)}_{xy}[S^2]}=i^{n_-(Q)-n_-(R)}\frac{\tilde{d}_x(Q)\tilde{d}_y(R)}{\tilde{d}_y(Q)\tilde{d}_x(R)}. \label{eq:QA-12}
\end{align}
Meanwhile, the relation $\tilde{\Theta}_x=\tilde{\Theta}_yC_{2z}$ leads to
\begin{align}
\tilde{d}_{C_{2z}}(\bm{k})=s\tilde{d}_y(C_{2z}\bm{k})\tilde{d}_x(\bm{k})^*, \label{eq:QA-13}
\end{align}
with $s=\pm 1$. Here, $\tilde{d}_{C_{2z}}$ is defined on the pass $l_0(t)=(\pi+\frac{r}{2}\sin t, \pi, \pi+\frac{r}{2}(1+\cos t))\ (0\le t\le\pi)$.
Moreover, as we show below, we have
\begin{align}
    \tilde{d}_{C_{2z}}(R)=i^{-C[S^2_+]}\tilde{d}_{C_{2z}}(Q). \label{eq:QA-14}
\end{align}
Then, combining Eqs.~(\ref{eq:QA-12})-(\ref{eq:QA-14}), we obtain Eq.~(\ref{eq:Q3-9}).

Next, to complete the proof,  we give a proof of Eq.~(\ref{eq:QA-14}). Due to $C_{2z}$ symmetry, we have
\begin{align}
    C[S^2_+]&=\frac{1}{\pi}\int_{(S^2_+)^{(0)}}d\bm{S}\cdot\mathrm{rot}\bm{A}(\bm{k}) \notag \\
    &=\frac{1}{\pi}\int_{l_0}d\bm{k}\cdot[\bm{A}(\bm{k})-\bm{A}(C_{2z}\bm{k})]. \label{eq:QA-15}
\end{align}
Here, $(S^2_+)^{(0)}$ is a half part of $S^2_+$: $(S^2_+)^{(0)}=\{\bm{k}=(\pi+\frac{r}{2}\sin\theta\cos\phi, \pi+\frac{r}{2}\sin\theta\sin\phi, \pi+\frac{r}{2}(1+\cos\theta))|\ 0\le\theta\le\pi,\ 0\le\phi\le \pi \}$, and $\bm{A}(\bm{k})$ is the Berry connection calculated by using the smooth gauge on $(S^2_+)^{(0)}$.
Meanwhile, one can easily find
\begin{align}
    A(C_{2z}\bm{k})=A(\bm{k})-i\nabla_{\bm{k}}\ln\det[\omega_{C_{2z}}(\bm{k})]. \label{eq:QA-16}
\end{align}
Moreover, from the definition of $\tilde{d}_{C_{2z}}$, we have
\begin{align}
    \frac{\tilde{d}_{C_{2z}}(R)}{\tilde{d}_{C_{2z}}(Q)}=\exp\qty[\frac{1}{2}\int_{l_0}d\bm{k}\cdot\nabla_{\bm{k}}\ln\det\omega_{C_{2z}}(\bm{k})]. \label{eq:QA-17}
\end{align}
Then, combining Eqs.~(\ref{eq:QA-15})-(\ref{eq:QA-17}), we obtain Eq.~(\ref{eq:QA-14}). 

Next,  Eq.~(\ref{eq:Q3-9-2}) is obtained by using Eqs.~(\ref{eq:Q3-7}) and (\ref{eq:Q3-9}).

\section{\label{sec:QB} Proof of some properties of $\mathcal{Q}_{xy}[T]$ }
In this section, we prove some properties of $\mathcal{Q}_{xy}[T]$ given in the main text. To this end, we use the parametrization $(t,k)\in[-\pi,\pi]^2$ of $T$ given in Sec.~\ref{sec:Q3-2}. Moreover, in the same manner as $\mathcal{Q}_{xy}[S^2]$, we rewrite the definition of $\mathcal{Q}_{xy}[T]$ as
\begin{align}
    (-1)^{\mathcal{Q}_{xy}[T]}=\prod_{i=0,1}\frac{\mathrm{Pf}[\omega_{x_i}(K_i)]}{\tilde{d}_{x_i}(K_i)}\frac{\mathrm{Pf}[\omega_{x_i}(K'_i)]}{\tilde{d}_{x_i}(K_i')}, \label{eq:QB-1}
\end{align}
where $\tilde{d}_{x_i}: [-\pi,\pi]^2\rightarrow S^1$ is a lift of $d_{x_i}: [-\pi,\pi]^2\rightarrow S^1\ (\ (t,k)\mapsto\mathrm{det}[\omega_{x_i}(\bm{k}(t,k))]\ )$ through the projection $p:S^1\rightarrow S^1\ (z\mapsto z^2)$. We note that $\tilde{d}_{x_i}$ is periodic along the $t$- and $k$- directions because one can show that $(d_{x_i})_*[\pi_1(T)]\subset \pi_1(S^1)$ is the trivial subgroup~\cite{Hatcher2001}.

Below, $\tilde{d}_G: [-\pi,\pi]^2\rightarrow S^1$ denotes a lift of $d_G:[-\pi,\pi]^2\rightarrow S^1\ (\ (t,k)\mapsto\mathrm{det}[\omega_{G}(\bm{k}(t,k))]\ )$ through the projection $p$, where $G$ is a unitary operator or an antiunitary operator and $[\omega_G(\bm{k})]_{mn}=\mel{u_{m}(G\bm{k})}{G}{u_n(\bm{k})}\ (m,n=1,2,\dots N_{\mathrm{occ}})$. Here, we note that $\tilde{d}_G$ is not always periodic along the $t$- and $k$-directions, because $T$ is not simply connected.
Moreover, $N_{G}\in\mathbb{Z}$ denotes the winding number of $d_G\circ l_t: [-\pi,\pi]\rightarrow S^1$, with $l_t(k)=(t,k)\ (-\pi\le k\le\pi)$. We note that $N_G$ does not depend on $t$ when we fix a continuous gauge because $d_G\circ l_t$ is homotopic to $d_G\circ l_{t'}$ for $t,t'\in[-\pi,\pi]$. We also note that $N_G$ is equal to the winding number of $\omega_G\circ l_t: [-\pi,\pi]\rightarrow U(N_{\mathrm{occ}})$.

\subsection{\label{sec:QB-1}Well-definedness of $\mathcal{Q}_{xy}[T]$}
In this subsection, we show that $\mathcal{Q}_{xy}[T]$ is well-defined, as claimed in Sec.~\ref{sec:Q3-2}.
The proof is almost the same as that for $\mathcal{Q}_{xy}[S^2]$. Here we only show that the value of $\mathcal{Q}_{xy}[T]$ does not depend on the choice of the quarter part of the torus $T$, e.g., we can use the pair $(K_2,K_2')$ instead of $(K_0,K_0')$ in Eq.~(\ref{eq:QB-1}). Firstly, we have 
\begin{align}
    \tilde{d}_x(\pi,k)=se^{i\frac{N_{\mathrm{occ}}}{2}r}\tilde{d}_{C_{2z}}(0,-k)\tilde{d}_{C_{2z}}(0,k)\tilde{d}_x(0,k), \label{eq:QB-12}
\end{align}
corresponding to Eq.~(\ref{eq:QA-7}), and 
\begin{align}
    &\mathrm{Pf}[\omega_x(K_2^{(')})] \notag \\
    &=(-1)^{\frac{N_{\mathrm{occ}}}{2}f}e^{i\frac{N_{\mathrm{occ}}}{2}r}\mathrm{det}[\omega_{C_{2z}}(K_0^{(')})]\mathrm{Pf}[\omega_x(K_0^{(')})], \label{eq:QB-13}
\end{align}
corresponding to Eq.~(\ref{eq:QA-8}).
Next, one can see $\tilde{d}_{C_{2z}}(K_0'')=\tilde{d}_{C_{2z}}(K_0')$ from $N_{C_{2z}}=0$. Then, using these equations, we obtain
\begin{align}
   (-1)^{\mathcal{Q}_{xy}[T]}=\prod_{i=1,2}\frac{\mathrm{Pf}[\omega_{x_i}(K_i)]}{\tilde{d}_{x_i}(K_i)}\frac{\mathrm{Pf}[\omega_{x_i}(K'_i)]}{\tilde{d}_{x_i}(K_i')}. \label{eq:QB-14}
\end{align}
Thus, by comparing with Eq.~(\ref{eq:QB-1}), we see that whether we use $(K_0,K_0')$ or $(K_2,K_2')$ does not affect the value of $\mathcal{Q}_{xy}[T]$.

\subsection{\label{sec:QB-3}Proof of Eq.~(\ref{eq:Q5-1-1})}
In this subsection, we give a proof of Eq.~(\ref{eq:Q5-1-1}) in Sec.~\ref{sec:Q5-1}.
Here, we focus on spinful MSG \#50.283. A similar discussion applies to other MSGs.
Firstly, using the relations $\tilde{\Theta}_x\mathcal{PT}=\{E|010\}\mathcal{PT}\tilde{\Theta}_x$ and $\tilde{\Theta}_y\mathcal{PT}=\{E|100\}\mathcal{PT}\tilde{\Theta}_y$, we have
\begin{align}
\omega_{\mathcal{PT}}(0,-k)&=-\omega_x(0,k)\omega_{\mathcal{PT}}(0,k)\omega_x(0,k)^T, \\
\omega_{\mathcal{PT}}(\frac{\pi}{2},-k)&=-\omega_y(\frac{\pi}{2},k)\omega_{\mathcal{PT}}(\frac{\pi}{2},k)\omega_x(\frac{\pi}{2},k)^T.
\end{align}
Next, since $(\mathcal{PT})^2=-1$, one can define $\mathrm{Pf}[\omega_{\mathcal{PT}}(t,k)]$ for all $(t,k)$, and we have
\begin{align}
    \tilde{d}_x(0,k)&=i^{\frac{N_{\mathrm{occ}}}{2}}\tilde{p}_{\mathcal{PT}}(0,k)\tilde{p}_{\mathcal{PT}}(0,-k), \label{eq:QB-21} \\
       \tilde{d}_y(\frac{\pi}{2},k)&=i^{\frac{N_{\mathrm{occ}}}{2}}\tilde{p}_{\mathcal{PT}}(\frac{\pi}{2},k)\tilde{p}_{\mathcal{PT}}(\frac{\pi}{2},-k), \label{eq:QB-22}
\end{align}
where $\tilde{p}_{\mathcal{PT}}(t,k)$ is a lift of $\mathrm{Pf}[\omega_{\mathcal{PT}}(t,k)]$ through $p$. 
Moreover, since the winding number of $\mathrm{Pf}[\omega_{\mathcal{PT}}]\circ l_t$ is equal to $N_\mathcal{PT}/2$, we also have
\begin{align}
    (-1)^{\frac{N_{\mathcal{PT}}}{2}}=\frac{\tilde{p}_{\mathcal{PT}}(K_0')}{\tilde{p}_{\mathcal{PT}}(K_0'')}=\frac{\tilde{p}_{\mathcal{PT}}(K_1')}{\tilde{p}_{\mathcal{PT}}(K_1'')}. \label{eq:QB-23}
\end{align}
Then, combining Eqs.~(\ref{eq:QB-21})-(\ref{eq:QB-23}), we obtain
\begin{align}
    (-1)^{\mathcal{Q}_{xy}[T]}=\prod_{i=0,1}\frac{\mathrm{Pf}[\omega_{x_i}(K_i)]}{\mathrm{Pf}[\omega_{\mathcal{PT}}(K_i)]}\frac{\mathrm{Pf}[\omega_{\mathcal{PT}}(K_i')]}{\mathrm{Pf}[\omega_{x_i}(K_i')]}. \label{eq:QB-24}
\end{align}
Meanwhile, decomposing the occupied bands at $K_0^{(')}$ into the direct sum of two-dimensional irreducible co-representations distinguished by the eigenvalue of $C_{2x}=(-1)^f\tilde{\Theta}_x\mathcal{PT}$, one can see
\begin{align}
    \mathrm{Pf}[\omega_x(K_0^{(')})]=\mathrm{Pf}[\omega_{\mathcal{PT}}(K_0^{(')})]\prod_{i=1}^{N_{\mathrm{occ}}/2}\zeta_{2i}[K_0^{(')}]. \label{eq:QB-25}
\end{align}
Similarly, we have
\begin{align}
    \mathrm{Pf}[\omega_y(K_1^{(')})]=\mathrm{Pf}[\omega_{\mathcal{PT}}(K_1^{(')})]\prod_{i=1}^{N_{\mathrm{occ}}/2}\xi_{2i}[K_1^{(')}]. \label{eq:QB-26}
\end{align}
Finally, combining Eqs.~(\ref{eq:QB-24})-(\ref{eq:QB-26}), we have Eq.~(\ref{eq:Q5-1-1}).

\subsection{\label{sec:QB-5}Proof of Eq.~(\ref{eq:Q5-1-2})}
In this subsection, we give a proof of Eq.~(\ref{eq:Q5-1-2}) in Sec.~\ref{sec:Q5-1}.
Firstly, using the relations $\tilde{\Theta}_x\tilde{C}_{2y}\mathcal{T}=(-1)^f\tilde{C}_{2y}\mathcal{T}\tilde{\Theta}_x$ and $\tilde{\Theta}_y\tilde{C}_{2x}\mathcal{T}=(-1)^f\tilde{C}_{2x}\mathcal{T}\tilde{\Theta}_y$, and noticing that $(\tilde{C}_x\mathcal{T})^2=-1$ on the plane $k_y=\pi$ and $(\tilde{C}_y\mathcal{T})^2=-1$ on the plane $k_x=\pi$, one can see
\begin{align}
 \tilde{d}_x(0,k)&=i^{\frac{N_{\mathrm{occ}}}{2}f}\tilde{p}_{\tilde{C}_{2y}\mathcal{T}}(0,k)\tilde{p}_{\tilde{C}_{2y}\mathcal{T}}(0,-k), \label{eq:QB-31} \\
 \tilde{d}_y(\frac{\pi}{2},k)&=i^{\frac{N_{\mathrm{occ}}}{2}f}\tilde{p}_{\tilde{C}_{2x}\mathcal{T}}(\frac{\pi}{2},k)\tilde{p}_{\tilde{C}_{2x}\mathcal{T}}(\frac{\pi}{2},-k). \label{eq:QB-32}
\end{align}
Then, substituting Eqs.~(\ref{eq:QB-31}) and (\ref{eq:QB-32}) into Eq.~(\ref{eq:QB-1}), we have
\begin{align}
        &(-1)^{\mathcal{Q}_{xy}[T]+(N_{\tilde{C}_{2x}\mathcal{T}}+N_{\tilde{C}_{2y}\mathcal{T}})/2} \notag \\
        &=\prod_{i=0,1}\frac{\mathrm{Pf}[\omega_{x_i}(K_i)]}{\mathrm{Pf}[\omega_{\tilde{C}_{2x_{i+1}}\mathcal{T}}(K_i)]}\frac{\mathrm{Pf}[\omega_{\tilde{C}_{2x_{i+1}}\mathcal{T}}(K_i')]}{\mathrm{Pf}[\omega_{x_i}(K'_i)]}. \label{eq:QB-33}
\end{align}
Meanwhile, using $\tilde{C}_{2x}\mathcal{T}\cdot \tilde{C}_{2y}\mathcal{T}=(-1)^f\{E|1\bar{1}0\}\tilde{C}_{2y}\mathcal{T}\cdot \tilde{C}_{2x}\mathcal{T}$, we find 
\begin{align}
    N_{\tilde{C}_{2x}\mathcal{T}}=N_{\tilde{C}_{2y}\mathcal{T}}\ (\ \in 2\mathbb{Z}\ ). \label{eq:QB-34}
\end{align}
Moreover, similar to the previous paragraph, one can show
\begin{align}
 &\mathrm{Pf}[\omega_x(K_0^{(')})] \notag \\
 &=(-1)^{\frac{N_{\mathrm{occ}}}{2}f}\mathrm{Pf}[\omega_{\tilde{C}_{2y}\mathcal{T}}(K_0^{(')})]\prod_{i=1}^{N_{\mathrm{occ}}/2}\zeta_{2i}[K_0^{(')}], \label{eq:QB-35} \\
  &\mathrm{Pf}[\omega_y(K_1^{(')})] \notag \\
  &=\mathrm{Pf}[\omega_{\tilde{C}_{2x}\mathcal{T}}(K_1^{(')})]\prod_{i=1}^{N_{\mathrm{occ}}/2}\zeta_{2i}[K_1^{(')}], \label{eq:QB-36}
\end{align}
where $\zeta_i(\bm{k})$ is the $M_z=\{M_{001}|0\}$ eigenvalue of the $i$-th occupied band at $\bm{k}$. Here, since the points on the planes $k_z=0$ and $k_z=\pi$ are $M_z$-invariant, $\zeta_{i}[K_0^{(')}]=\zeta_{i}[K_1^{(')}]$. Then, combining Eqs.~(\ref{eq:QB-33})-(\ref{eq:QB-36}), we finally obtain $(-1)^{\mathcal{Q}_{xy}[T]}=1$.

\subsection{\label{sec:QB-6}Proof of Eqs.~(\ref{eq:Q5-2-1})-(\ref{eq:Q5-2-4})}
In this subsection, we give proofs of Eqs.~(\ref{eq:Q5-2-1})-(\ref{eq:Q5-2-4}) in Sec.~\ref{sec:Q5-2}. To this end, we first show that, when the system has a symmetry $G$ which sends $(t=0,k)$ to $(t=\pm\pi/2, \pm k)$ and satisfies
\begin{align}
    \tilde{\Theta}_yG=(-1)^{s}\{E|ab0\}G\tilde{\Theta}_x, \label{eq:QB-61}
\end{align}
with $a,b\in\mathbb{N}$ and $s\in\{0,1\}$, then we have
\begin{align}
    \mathcal{Q}_{xy}[T]=N_G\ \ (\mathrm{mod}\ 2). \label{eq:QB-62}
\end{align}
Below, we assume that $G$ is unitary and sends $(0,k)$ to $(\pi/2,k)$. A similar discussion applies to other cases.
Firstly, using Eq.~(\ref{eq:QB-61}), we have 
\begin{align}
\mathrm{Pf}[\omega_y(K_1^{(')})] &=(-1)^{\frac{N_{\mathrm{occ}}}{2}s}A\cdot\mathrm{det}[\omega_G(K_0^{(')})]\mathrm{Pf}[\omega_x(K_0^{(')})], \\
\tilde{d_y}(\pi/2,k)&=A\cdot\tilde{d}_G(0,k)\tilde{d}_G(0,-k)\tilde{d}_x(0,k),
\end{align}
with $A=e^{-i\frac{N_{\mathrm{occ}}}{2}[a\pi+b(\pi+r)]}$. Next, substituting these equations into Eq.~(\ref{eq:QB-1}), one can show
\begin{align}
    (-1)^{\mathcal{Q}_{xy}[T]}&=\frac{\tilde{d}_G(K_0')}{\tilde{d}_G(K_0'')}.
\end{align}
Then, since $\tilde{d}_G(K_0')=(-1)^{N_G}\tilde{d}_G(K_0'')$, we finally obtain Eq.~(\ref{eq:QB-62}).

Next, we give a proof of Eq.~(\ref{eq:Q5-2-1}). To this end, we focus on systems with MSG \#100.175, and set $G=C_{4z}=\{4^+_{001}|0\}$. [Almost the same discussion holds for MSG \#100.173 with $G=C_{4z}\mathcal{T}$.] Firstly,  Eq.~(\ref{eq:QB-61}) holds with $a=1, b=0,$ and $s=0$, and thus we have Eq.~(\ref{eq:QB-62}). Next, using $C_{4z}^2=C_{2z}$, we have $2N_G=N_{C_{2z}}$. Here, as shown in the previous paragraph, $N_{C_{2z}}=0$. Therefore, we finally obtain $\mathcal{Q}_{xy}[T]=N_{C_{2z}}/2=0\ (\mathrm{mod}\ 2)$. 

Next, we give a proof of Eq.~(\ref{eq:Q5-2-2}). To this end, we focus on systems with MSG \#106.223, and set $G=\tilde{C}_{4,2}=\{4^+_{001}|00\frac{1}{2}\}$. [Almost the same discussion holds for MSG \#106.221 with $G=\tilde{C}_{4,2}\mathcal{T}$.] Firstly,  Eq.~(\ref{eq:QB-61}) holds with $a=1, b=0,$ and $s=0$, and thus we have Eq.~(\ref{eq:QB-62}). Next, using $(\tilde{C}_{4,2})^2=\{E|001\}C_{2z}$, we have $2N_G=N_{e}$ with $e(k)=e^{-iN_{\mathrm{occ}}k}\ (-\pi\le k\le\pi)$. Meanwhile, one can easily see $N_{e}=-N_{\mathrm{occ}}$. Therefore, we finally obtain $\mathcal{Q}_{xy}[T]=N_G=N_{\mathrm{occ}}/2\ (\mathrm{mod}\ 2)$.

Next, we give a proof of Eq.~(\ref{eq:Q5-2-3}). To this end, we focus on systems with MSG \#117.301, and set $G=C_2=\{2_{110}|\frac{1}{2} \bar{\frac{1}{2}} 0\}$. Firstly,  Eq.~(\ref{eq:QB-61}) holds with $a=0, b=1,$ and $s=0$, and thus we have Eq.~(\ref{eq:QB-62}). Next, we define $\phi:[-\pi,\pi]\rightarrow S^1$ as $\phi(k)=d_{G}(l_{\pi/4}(k))d_{G}(l_{\pi/4}(0))^{-1}$. Then, using $\phi(-k)\phi(k)=1$ and $\phi(0)=1$, we have $\tilde{\phi}(-k)=\tilde{\phi}(k)^*$, where $\tilde{\phi}:[-\pi,\pi]\rightarrow S^2$ is a lift of $\phi$ through $p:S^1\rightarrow S^1$. Therefore, we find
\begin{align}
    &\ N_G=0\ (\mathrm{mod}\ 2) \notag \\
    \Leftrightarrow&\ \tilde{\phi}(-\pi)=\tilde{\phi}(\pi) \notag \\
    \Leftrightarrow&\ \tilde{\phi}(\pi)\in\mathbb{R} \notag \\
    \Leftrightarrow&\ d_{G}(\frac{\pi}{4}, \pi)=d_{G}(\frac{\pi}{4},0).
\end{align}
Then we also find that $N_G=1\ (\mathrm{mod}\ 2)$ if and only if $d_{G}(\pi/4,\pi)=-d_{G}(\pi/4,0)$. These relations are summarized as $(-1)^{\mathcal{Q}_{xy}[T]}=(-1)^{N_G}=\det \omega_G(S)/\det \omega_G(\Sigma)$, which corresponds to Eq.~(\ref{eq:Q5-2-3}).

Finally, we give a proof of Eq.~(\ref{eq:Q5-2-4}). To this end, we focus on systems with MSG \#117.303, and set $G=C_2\mathcal{T}=\{2_{1\bar{1}0}|\frac{1}{2}\frac{1}{2}0\}'$. Firstly,  Eq.~(\ref{eq:QB-61}) holds with $a=1, b=0,$ and $s=f$, and thus we have Eq.~(\ref{eq:QB-62}).
Next, when there are no gapless nodes on the plane $k_x=k_y$ outside of $T$, we can extend the continuous gauge on $T$ to the region $\{\bm{k}| k_x=k_y,\ 0\le k_x\le\pi-r/\sqrt{2},\ -\pi\le k_z\le \pi\}$. Then we find $\mathcal{Q}_{xy}[T]=N_G=N_{G}[l]\ (\mathrm{mod}\ 2)$ with $l(k)=(0,0,k)\ (-\pi\le k\le\pi)$. Next, using $C_2\mathcal{T}\bar{C_4}=(-1)^f\{E|010\}\bar{C_4}C_2\mathcal{T}C_{2z}$, one can show $N_{G}[l]=-N_{\bar{c_4}}[l]$. Then, using $\det \omega_{\bar{C_4}}(0,0,-k)\det \omega_{\bar{C_4}}(0,0,k)=\det \omega_{C_{2z}}(0,0,k)$ and noticing that $\det \omega_{C_{2z}}(0,0,k)$ is constant on the line $l$, we have $(-1)^{\mathcal{Q}_{xy}[T]}=\det \omega_{\bar{C_4}}(Z)/\det \omega_{\bar{C_4}}(\Gamma)$, which corresponds to Eq.~(\ref{eq:Q5-2-4}).

\section{\label{sec:QC}Definitions of $H(t,k)$ and $\tilde{H}(s,k)$}
In this section, we give the definitions of $H(t,k)$ and $\tilde{H}(s,k)$ used in Sec.~\ref{sec:Q4}. Below, $H(\bm{k})\ (\ \bm{k}=(k_x,k_y,k_z)\in(\mathbb{R}/2\pi\mathbb{Z})^3 \ )$ denotes the Bloch Hamiltonian of the original 3D system with MSG \#32.138, which satisfies
\begin{align}
\tilde{\Theta}_x(k_x,k_y)H(\bm{k})\tilde{\Theta}_x(k_x,k_y)^{-1}&=H(k_x,-k_y,-k_z), \label{eq:QC-1} \\
\tilde{\Theta}_y(k_x,k_y)H(\bm{k})\tilde{\Theta}_y(k_x,k_y)^{-1}&=H(-k_x,k_y,-k_z),  \\
\tilde{\Theta}_x(k_x,-k_y)\tilde{\Theta}_x(k_x,k_y)&=e^{-ik_y}, \\
\tilde{\Theta}_y(-k_x,k_y)\tilde{\Theta}_y(k_x,k_y)&=e^{-ik_x}, \label{eq:QC-2}
\end{align}
where $\tilde{\Theta}_{x_i}(k_x,k_y)\ (x_i=x,y)$ represents the antiunitary operation $\tilde{\Theta}_{x_i}=\mathcal{G}_{x_i}\mathcal{T}$ under the Bloch basis set.

Firstly, we define $H(t,k)\ (\ (t,k)\in(\mathbb{R}/2\pi\mathbb{Z})^2\ )$ as the 2D subsystem on $T$ obtained by restricting the original 3D system $H(\bm{k})$:
\begin{align}
 H(t,k)=H(\pi+r\cos t, \pi+r\sin t, k). 
\end{align}
Since the original system $H(\bm{k})$ is assumed to be gapped on $T$, the 2D system $H(t,k)$ is also gapped. Moreover, from Eqs.~(\ref{eq:QC-1})-(\ref{eq:QC-2}), we obtain 
\begin{align}
\tilde{\Theta}_x(t)H(t,k)\tilde{\Theta}_x(t)^{-1}&=H(-t,-k), \label{eq:QC-3}\\
\tilde{\Theta}_y(t)H(t,k)\tilde{\Theta}_y(t)^{-1}&=H(\pi-t,-k), \label{eq:QC-4}\\
\tilde{\Theta}_x(-t)\tilde{\Theta}_x(t)&=-e^{-ir\sin t}, \label{eq:QC-5}\\ 
\tilde{\Theta}_y(\pi-t)\tilde{\Theta}_y(t)&=-e^{-ir\cos t}, \label{eq:QC-6}
\end{align}
with $\tilde{\Theta}_x(t)=\tilde{\Theta}_x(\pi+r\cos t, \pi+r\sin t)$ and $\tilde{\Theta}_y(t)=\tilde{\Theta}_y(\pi+r\cos t, \pi+r\sin t)$. 

Next, we define $\tilde{H}(s,k)\ (\ (s,k)\in(\mathbb{R}/2\pi\mathbb{Z})^2\ )$ so that $H(t,k)\ (0\le t\le\frac{\pi}{2})$ and $\tilde{H}(2t,k)$ have the same bulk/edge spectra, and it preserves an antiunitary symmetry $\tilde{\Theta}$ with $\tilde{\Theta}^2=-1$. Briefly, we identify the $\tilde{\Theta}_x$ symmetry acting on the line $t=0$ and the $\tilde{\Theta}_y$ symmetry acting on the line $t=\pi/2$ by a unitary transformation, and extend the identified symmetry $\tilde{\Theta}$ to the whole BZ. The detailed procedure is as follows. First, from Eq.~(\ref{eq:QC-5}), one can find a smooth unitary transformation $U_x: (-\delta_x,\delta_x)\rightarrow U(2N)$ $(0<\delta_x<\frac{\pi}{6})$ satisfying
\begin{align}
U_x(-t)^{\dagger}e^{-i\frac{r\sin t}{2}}\tilde{\Theta}_x(t)U_x(t)=\tilde{\Theta}, \label{eq:QC-7}
\end{align}
where $\tilde{\Theta}=i(\sigma_2\otimes I_{N})K$ and $N$ is half of the number of sites in the unit cell (see below). Similarly, from Eq.~(\ref{eq:QC-6}), one can find another smooth unitary transformation $U_y: (\frac{\pi}{2}-\delta_y,\frac{\pi}{2}+\delta_y)\rightarrow U(2N)$ $(0<\delta_y<\frac{\pi}{6})$ satisfying
\begin{align}
    U_y(\pi-t)^{\dagger}e^{-i\frac{r\cos t}{2}}\tilde{\Theta}_y(t)U_y(t)=\tilde{\Theta}. \label{eq:QC-8}
\end{align}
Next, we take a smooth unitary transformation $U:[0,\frac{\pi}{2}]\rightarrow U(2N)$ satisfying $U(t)=U_x(t)\ (0\le t\le \frac{\delta_x}{2})$ and $U(t)=U_y(t)\ (\frac{\pi}{2}-\frac{\delta_y}{2}\le s\le \frac{\pi}{2})$. Next, by using the transformation, we define 
\begin{align}
    \tilde{H}^{>}(s,k)&=U(s/2)^{\dagger}H(s/2,k)U(s/2) \ \ (0\le s\le \pi), \label{eq:QC-9} \\
    \tilde{H}^{<}(s,k)&=\tilde{\Theta}\tilde{H}^>(-s,-k)\tilde{\Theta}^{-1}\ \ (-\pi\le s\le 0). \label{eq:QC-10}
\end{align}
Then, one can easily see $\tilde{H}^>(0,k)=\tilde{H}^<(0,k)$ and $\tilde{H}^>(\pi,k)=\tilde{H}^<(-\pi,k)$. Therefore, we can define
\begin{align}
    \tilde{H}(s,k)=
     \begin{cases}
    \tilde{H}^>(s,k)   & \text{($0\le s\le\pi$),} \\
    \tilde{H}^<(s,k)   & \text{($-\pi\le s\le 0$),} 
     \end{cases}
\end{align}
for $(s,k)\in(\mathbb{R}/2\pi\mathbb{Z})^2$. It satisfies 
\begin{align}
    \tilde{\Theta}\tilde{H}(s,k)\tilde{\Theta}^{-1}&=H(-s,-k),
\end{align}
and is smooth with respect to $s$ and $k$. In fact, it is obvious that $\tilde{H}(s,k)$ is smooth except for the lines $s=0,\pi$. $\tilde{H}(s,k)$ is also smooth on the lines because $\tilde{H}(s,k)=U_x(s/2)^{\dagger}H(s/2,k)U_x(s/2)$ for $-\frac{\delta_x}{2}\le s\le\frac{\delta_x}{2}$, and $\tilde{H}(s,k)=U_y(s/2)^{\dagger}H(s/2,k)U_y(s/2)$ for $\frac{\pi}{2}-\frac{\delta_y}{2}\le s\le\frac{\pi}{2}+\frac{\delta_y}{2}$. Moreover, since $H(s/2,k)$ and $\tilde{H}(s,k)\ (0\le s\le\pi)$ are related by a unitary matrix $U(s/2)$, which does not depend on $k$, $H(s/2,k)$ and $\tilde{H}(s,k)\ (0\le t\le\pi)$ have same bulk spectra and edge spectra projected along the $k$-direction.

To complete the construction of $\tilde{H}(s,k)$, we show that, for antiunitary operator $\Theta(t)$ satisfying
\begin{align}
    \Theta(-t)\Theta(t)=-1,
\end{align}
there exists a smooth unitary transformation $U(t):(-\delta,\delta)\rightarrow U(2N)$ such that
\begin{align}
    U(-t)^{\dagger}\Theta(t)U(t)=\tilde{\Theta}. \label{eq:QC-11}
\end{align}
[It is used to obtain Eqs.~(\ref{eq:QC-7}) and (\ref{eq:QC-8}).] To this end, firstly, for a given vector $u(t)\in\mathbb{C}^{2N}$ with $|u(t)|=1$, we define other vectors $u^{\alpha/\beta}(t)\in\mathbb{C}^{2N}$ as
\begin{align}
    u^{\alpha}(t)&=\frac{u(t)-b^*(t)\Theta(-t)u(-t)}{|u(t)-b^*(t)\Theta(-t)u(-t)|}, \\
    u^{\beta}(t)&=\frac{\Theta(-t)u(-t)-\beta(t)u(t)}{|\Theta(-t)u(-t)-\beta(t)u(t)|},
\end{align}
where $b(t)=a(t)/[1+\sqrt{1-|a(t)|^2}]$, and $a(t)=(u(t),\Theta(-t)u(-t))$. Then, one can see that these vectors satisfy the equations
\begin{align}
    (u^{\alpha}(t), u^{\beta}(t))&=0, \label{eq:QC-12}\\
    \Theta(t)u^{\alpha}(t)&=u^{\beta}(-t), \label{eq:QC-13}\\
    \Theta(t)u^{\beta}(t)&=-u^{\alpha}(-t).\label{eq:QC-14}
\end{align}
Next, we take vectors $u_i\in\mathbb{C}^{2N}\ (i=1,2,\dots N)$ such that 
$[u_1 \ \Theta(0)u_1\ \dots \ u_N \ \Theta(0)u_N]\in U(2N)$. Next, we inductively define $u_i(t)$ as $u_1(t)=u_1$ and
\begin{align}
    u_i(t)&=\frac{f_i(t)}{|f_i(t)|}, \\
    f_i(t)&=u_i-\sum_{j<i}\qty[(u^{\alpha}_j(t),u_i)u^{\alpha}_j(t)+(u^{\beta}_j(t),u_i)u^{\beta}_j(t)],
\end{align}
for $1<i\le N$. Then, from Eq.~(\ref{eq:QC-12}), one can see that
\begin{align}
    U(t)=[u_1^{\alpha}(t)\ u_1^{\beta}(t) \ \dots \ u_N^{\alpha}(t) \ u_N^{\beta}(t)]
\end{align}
belongs to $U(2N)$ and is smooth on a neighborhood of $t=0$. Moreover, from Eqs.~(\ref{eq:QC-13}) and (\ref{eq:QC-14}), this unitary matrix $U(t)$ satisfies Eq.~(\ref{eq:QC-11}).

\section{\label{sec:QD}Topological charges in systems on a body-centered lattice}
In this section, we discuss the topological charges $\mathcal{Q}_{xy}[S^2(R)]$, $\mathcal{Q}_{xy}[S^2(S)]$, and $\mathcal{Q}_{xy}[T]$ in systems on a body-centered lattice, which are partially discussed in Sec.~\ref{sec:Q5-3}. We first derive explicit forms of the topological charges written in terms of MSG \#45.238. Next, we show that the charges can be written in simple forms when we add additional symmetries relating $\mathcal{GT}$-invariant HSPs.

\subsection{\label{sec:QD-1}Derivation of topological charge formulas in systems with MSG \#45.238 ($Ib'a'2$)}
Since \#45.238 is a supergroup of \#32.138, the topological charges $\mathcal{Q}_{xy}[S^2(R)]$, $\mathcal{Q}_{xy}[S^2(S)]$, and $\mathcal{Q}_{xy}[T]$ defined in Sec.~\ref{sec:Q3} are also defined in systems with \#45.238. The explicit forms of the charges written in terms of \#45.238 are derived as follows.

Firstly, we double the unit cell of the system to reduce the symmetry of the system from \#45.236 to \#32.138. The doubled unit cell consists of two unit cells $U_1$ and $U_2$ of \#45.236 related by a transition vector $\bm{a}=(\frac{1}{2}, \frac{1}{2}, \frac{1}{2})$. Then, the Bloch Hamiltonian $H(\bm{k})$ and the antiunitary operator $\tilde{\Theta}_{x_i}(\bm{k})\ (x_i=x,y)$ of a given system with MSG \#45.238 are rewritten as~\cite{Kim2020}
\begin{align}
    H'(\bm{k})&=\mqty( H_1(\bm{k}) & e^{-i\bm{k}\cdot \bm{a}}H_2(\bm{k}) \\ e^{i\bm{k}\cdot \bm{a}}H_2(\bm{k}) & H_1(\bm{k}) ), \\
    \tilde{\Theta}_{x_i}'(\bm{k})&=\mqty(\tilde{\Theta}_{x_i}(\bm{k}) & O \\ O& e^{ik_{x_i}
}\tilde{\Theta}_{x_i}(\bm{k})),
\end{align}
where $H_1(\bm{k})=[H(\bm{k})+H(\bm{k+b})]/2$ describes intra-unit-cell hoppings,  $H_2(\bm{k})=[H(\bm{k})-H(\bm{k+b})]/2$ describes inter-unit-cell hoppings between $U_1$ and $U_2$, and $\bm{b}=(0,0,2\pi)$. One can easily see that  $H'(\bm{k})$ and $\tilde{\Theta}_{x_i}'(\bm{k})$ are periodic on the BZ for \#32.138, and satisfies
\begin{align}
H'(\tilde{\Theta}_{x_i}\bm{k})=\tilde{\Theta}_{x_i}'(\bm{k})H'(\bm{k})\tilde{\Theta}_{x_i}'(\bm{k})^{-1}.
\end{align}
[Here, we assumed that $\tilde{\Theta}_{x_i}(\bm{k})$ does not depend on $k_z$.] Therefore, the topological charges for \#32.138 are also defined for systems with MSG \#45.238 by using $H'(\bm{k})$ and $\tilde{\Theta}_{x_i}'(\bm{k})$. Then, since $H'(\bm{k})$ is constructed from $H(\bm{k})$, we can describe the topological charges in terms of the eigenvectors of $H(\bm{k})$ as desired. Below, we give detailed discussions for each of $\mathcal{Q}_{xy}[S^2(R)]$, $\mathcal{Q}_{xy}[S^2(S)]$, and $\mathcal{Q}_{xy}[T]$.

\begin{figure*}
\includegraphics[width=2\columnwidth, pagebox=cropbox, clip]{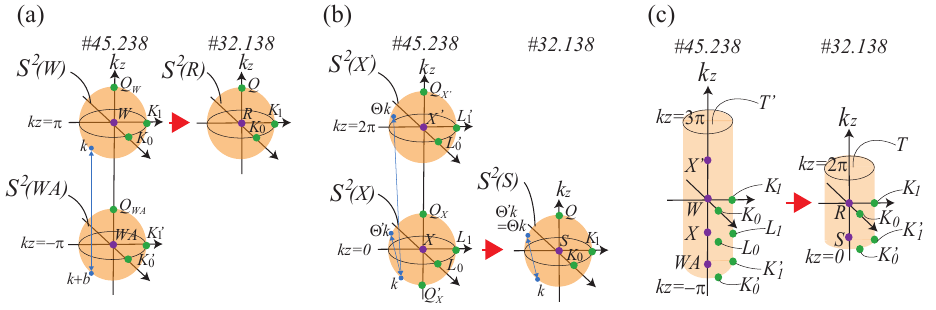}
\caption{\label{fig:QE}\small{Calculation of topological charges in MSG \#45.238 ($Ib'a'2$). (a) $Z_2\times Z_2$ charge $\mathcal{Q}_{xy}[S^2(R)]$.   (b) $Z_2\times Z_2$ charge $\mathcal{Q}_{xy}[S^2(S)]$. (c) $Z_2$ charge $\mathcal{Q}_{xy}[T]$. }}
\end{figure*}

\subsubsection{$Z_2\times Z_2$ charge $\mathcal{Q}_{xy}[S^2(R)]$}
We first discuss the formula of $\mathcal{Q}_{xy}[S^2(R)]$ in \#45.238. The point $R:(\pi,\pi,\pi)$ in the BZ for \#32.138 comes from two points  $W:(\pi,\pi,\pi)$ and $WA:(\pi,\pi,-\pi)$ in the BZ for \#45.238 (see Fig.~\ref{fig:QE}(a)). Here, the points $W$ and $WA$ are $\mathcal{GT}$-invariant HSPs. Therefore, we can define the $Z_2\times Z_2$ charges  $\mathcal{Q}_{xy}[S^2(W)]$ and $\mathcal{Q}_{xy}[S^2(WA)]$ as Eqs.~(\ref{eq:Q3-0})-(\ref{eq:Q3-1}). Then, $\mathcal{Q}_{xy}[S^2(R)]$ is written by using the $Z_2\times Z_2$ charges in \#45.236 as
\begin{align}
    \mathcal{Q}^{(i)}_{xy}[S^2(R)]=\mathcal{Q}^{(i)}_{xy}[S^2(W)]+\mathcal{Q}^{(i)}_{xy}[S^2(WA)]\ \  (\mathrm{mod}\ 2), \label{eq:QD-0}
\end{align}
for $i=0,1$.

Equation~(\ref{eq:QD-0}) is shown as follows. Firstly, let us take a continuous gauge of $H(\bm{k})$ on $S^2(W)$ and $S^2(WA)$, denoted by $\Phi(\bm{k})=[\ket{u_1(\bm{k})}\  \dots\ \ket{u_{N_{\mathrm{occ}}}(\bm{k})}]$. Then, we can choose a continuous gauge of $H'(\bm{k})$ on $S^2(R)$ as~\cite{Kim2020}
\begin{align}
    \Phi'(\bm{k})=\mqty(\Phi(\bm{k}) & \Phi(\bm{k+b}) \\ e^{i\bm{k}\cdot\bm{a}}\Phi(\bm{k}) & -e^{i\bm{k}\cdot\bm{a}}\Phi(\bm{k+b})). \label{eq:QD-1}
\end{align}
Using this gauge, the sewing matrix $\omega_{x_i}'(\bm{k})=\Phi'(\tilde{\Theta}_{x_i}\bm{k})^{\dagger}\tilde{\Theta}'_{x_i}(\bm{k})\Phi'(\bm{k})$ with $\tilde{\Theta}_{x}\bm{k}=(k_x, 2\pi-k_y, 2\pi-k_z)$ and $\tilde{\Theta}_{y}\bm{k}=(2\pi-k_x, k_y, 2\pi-k_z)$ is calculated as
\begin{align}
    \omega_{x_i}'(\bm{k})=\mqty(\omega_{x_i}(\bm{k}) & O \\ O & \omega_{x_i}(\bm{k+b})). \label{eq:QD-2}
\end{align}
Equation~(\ref{eq:QD-2}) leads to
\begin{align}
    \mathrm{Pf}[\omega_{x_i}'(K_i)]&=\mathrm{Pf}[\omega_{x_i}(K_i)]\mathrm{Pf}[\omega_{x_i}(K_i')], \label{eq:QD-3}\\
    \tilde{d}_{x_i}'(\bm{k})&=\tilde{d}_{x_i}(\bm{k})\tilde{d}_{x_i}(\bm{k+b}), \label{eq:QD-4} \\
    n'_-(Q)&=n_-(Q_W)+n_-(Q_{WA}), \label{eq:QD-5}
\end{align}
with $K_0:(\pi+r,\pi,\pi)$, $K_0':(\pi+r,\pi,-\pi)$, $K_1:(\pi,\pi+r,\pi)$, and $K_1':(\pi,\pi+r,-\pi)$ (see Fig.~\ref{fig:QE}(a)).
Then, substituting Eqs.~(\ref{eq:QD-3})-(\ref{eq:QD-5}) into Eqs.~(\ref{eq:Q3-0}) and (\ref{eq:Q3-1}), we obtain Eq.~(\ref{eq:QD-0}).

\subsubsection{the $Z_2\times Z_2$ charge $\mathcal{Q}_{xy}[S^2(S)]$}
We next discuss the formula of $\mathcal{Q}_{xy}[S^2(S)]$ in \#45.238. The point $S:(\pi,\pi,0)$ in the BZ for \#32.138 comes from two points  $X:(\pi,\pi,0)$ and $X':(\pi,\pi,2\pi)$ in the BZ for \#45.238 (see Fig.~\ref{fig:QE}(b)). Here, in contrast to $W$ and $WA$, the points $X$ and $X'$ are not $\mathcal{GT}$-invariant. Therefore, we cannot define the $Z_2\times Z_2$ charges for the points. Meanwhile, we can also describe $\mathcal{Q}_{xy}[S^2(S)]$ in terms of MSG \#45.238 as
\begin{align}
    \mathcal{Q}_{xy}^{(0)}[S^2(S)]&=\frac{C[S^2(X)]-\Delta n_-}{2}\ \ (\mathrm{mod}\ 2), \label{eq:QD-6} \\
    \mathcal{Q}_{xy}^{(1)}[S^2(S)]&=\frac{C[S^2(X)]+\Delta n_-}{2}\ \ (\mathrm{mod}\ 2). \label{eq:QD-6-1}
\end{align}
Here, $S^2(X)$ is a sphere centered at $X:(\pi,\pi,0)$, and $\Delta n_-=n_-(Q_X)-n_-(Q_X')$ is the difference of the number of occupied bands with $C_{2z}=-i^f$ between the two $C_{2z}$-invariant points on the sphere (see Fig.~\ref{fig:QE}(b)).

Equations~(\ref{eq:QD-6}) and (\ref{eq:QD-6-1}) is shown as follows. Firstly, let us take a continuous gauge of $H(\bm{k})$ on the half part of $S^2(X)$ and $S^2(X')$ with $k_y\ge\pi$.  Then, we can choose a continuous gauge of $H'(\bm{k})$ on the half part of $S^2(S)$ with $k_y\ge\pi$ as Eq.~(\ref{eq:QD-1}). Using this gauge, the sewing matrix $\omega'_{x_i}(\bm{k})=\Phi'(\tilde{\Theta}'_{x_i}\bm{k})^{\dagger}\tilde{\Theta}_{x_i}'(\bm{k})\Phi'(\bm{k})$ with $\tilde{\Theta}_{x}'\bm{k}=(k_x, 2\pi-k_y, -k_z)$ and $\tilde{\Theta}'_{y}\bm{k}=(2\pi-k_x, k_y, -k_z)$ is calculated as 
\begin{align}
        \omega_{x_i}'(\bm{k})=\mqty(O & \omega_{x_i}(\bm{k+b})  \\ \omega_{x_i}(\bm{k}) & O). \label{eq:QD-7}
\end{align}
Then, using the relations $\omega_{x_i}(\tilde{\Theta}_{x_i}'\bm{k}+\bm{b})=\omega_{x_i}(\tilde{\Theta}_{x_i}\bm{k})=-\omega_{x_i}(\bm{k})^T$ on the plane $k_{x_{i+1}}=\pi$ and $n_-(Q_{X'})=n_-(Q_X')$, we have
\begin{align}
    \mathrm{Pf}[\omega_{x_i}'(K_i)]&=(-1)^{\frac{N_{\mathrm{occ}}(N_{\mathrm{occ}}+1)}{2}}d_{x_i}(L_i), \label{eq:QD-8}\\
    \tilde{d}_{x_i}'(\bm{k})&=\tilde{d}_{x_i}(\bm{k})\tilde{d}_{x_i}(\tilde{\Theta}'_{x_i}\bm{k}), \label{eq:QD-9}\\
    n'_-(Q)&=n_-(Q_X)+n_-(Q'_X), \label{eq:QD-10}
\end{align}
with $L_0:(\pi+r,\pi,0)$ and $L_1:(\pi,\pi+r,0)$ (see Fig.~\ref{fig:QE}(b)). 
Substituting Eqs.~(\ref{eq:QD-8})-(\ref{eq:QD-10}) into Eq.~(\ref{eq:Q3-0}), we obtain
\begin{align}
    &(-1)^{\mathcal{Q}^{(0)}_{xy}[S^2]} \notag \\
    &=i^{N_{\mathrm{occ}}f+n_-(Q_X)+n_-(Q'_X)}\frac{\tilde{d}_x(Q_X)\tilde{d}_x(Q_X')}{\tilde{d}_y(Q_X)\tilde{d}_y(Q_X')}. \label{eq:QD-11}
\end{align}
Meanwhile, in the same manner as Appendix~\ref{sec:QA-5}, one can show that $\tilde{d}_{C_{2z}}$, a lift of $d_{C_{2z}}$ through $p$, satisfies
\begin{align}
    \tilde{d}_{C_{2z}}(\bm{k})&=\tilde{d}_y(C_{2z}\bm{k})\tilde{d}_x(\bm{k})^*, \label{eq:QD-12} \\
    \tilde{d}_{C_{2z}}(Q_X')&=i^{-C[S^2(X)]}\tilde{d}_{C_{2z}}(Q_X). \label{eq:QD-13}
\end{align}
Then, combining Eqs.~(\ref{eq:QD-11})-(\ref{eq:QD-13}), we finally obtain Eq.~(\ref{eq:QD-6}). Equation~(\ref{eq:QD-6-1}) is obtained from Eqs.~(\ref{eq:Q3-7}) and (\ref{eq:QD-6}).

\subsubsection{$Z_2$ charge $\mathcal{Q}_{xy}[T]$}
Finally, we discuss the formula of the $Z_2$ charge $\mathcal{Q}_{xy}[T]$ in \#45.238. The torus $T$ in the BZ for \#32.138 comes from the torus $T'=\{\bm{k} | (k_x, k_y, k_z)=(\pi+r\cos t, \pi+r\sin t, k)  \ \ (-2\pi\le k\le 2\pi\ ,-\pi\le t\le\pi)\}$ in the BZ for \#45.238 (see Fig.~\ref{fig:QE}(c)). Here, we can define the $Z_2$ charge $\mathcal{Q}_{xy}[T']$ for the torus $T'$ as Eq.~(\ref{eq:Q4-1}) by using $\tilde{\Theta}_{x_i}$-invariant HSPs $K_i,K_i'\ (i=0,1)$ on the torus. Then, one can show that the $Z_2$ charge $\mathcal{Q}_{xy}[T]$ in \#32.138 and the $Z_2$ charge $\mathcal{Q}_{xy}[T']$ in \#45.238 are the same:
\begin{align}
    \mathcal{Q}_{xy}[T]=\mathcal{Q}_{xy}[T']\ \ (\mathrm{mod}\ 2). \label{eq:QD-14}
\end{align}
Therefore, we can determine whether QHSSs appear on the (001) surface by using $\mathcal{Q}_{xy}[T']$.

Equation~(\ref{eq:QD-14}) is shown as follows. Firstly, let us take a continuous gauge of $H(\bm{k})$ on the torus $T'$. Then, we can choose a continuous gauge of $H'(\bm{k})$ on $T$ as
\begin{align}
     \Phi'(\bm{k})=\mqty(\Phi(\bm{k}) & \Phi(\bm{k+b}) \\ e^{i\bm{k}\cdot\bm{a}}\Phi(\bm{k}) & -e^{i\bm{k}\cdot\bm{a}}\Phi(\bm{k+b}))O(k_z), \label{eq:QD-15}
\end{align}
for $k_z\in [0,2\pi]$, where $O(k_z): [0,2\pi]\rightarrow SO(2N_{\mathrm{occ}})$ satisfies $O(0)=I_{2N_{\mathrm{occ}}}$ and $O(2\pi)=\sigma_x\otimes I_{N_{\mathrm{occ}}}$. [Since $N_{\mathrm{occ}}\in 2\mathbb{Z}$, we have $\sigma_x\otimes I_{N_{\mathrm{occ}}}\in SO(2N_{\mathrm{occ}})$.] Using this gauge, the sewing matrix $\omega_{x_i}'(\bm{k})=\Phi'(\tilde{\Theta}_{x_i}\bm{k})^{\dagger}\tilde{\Theta}_{x_i}'(\bm{k})\Phi'(\bm{k})$ is calculated as
\begin{align}
    \omega_{x_i}'(\bm{k})=O(2\pi-k_z)^T\mqty(\omega_{x_i}(\bm{k}) & O \\ O & \omega_{x_i}(\bm{k+b}))O(k_z). \label{eq:QD-16}
\end{align}
Then we have
\begin{align}
    \mathrm{Pf}[\omega_{x_i}'(K_i)]&=\mathrm{Pf}[\omega_{x_i}(K_i)]\mathrm{Pf}[\omega_{x_i}(K_i')], \label{eq:QD-17}\\
    \mathrm{Pf}[\omega_{x_i}'(K'_i)]&=(-1)^{\frac{N_{\mathrm{occ}}(N_{\mathrm{occ}}+1)}{2}}d_{x_i}(L_i), \label{eq:QD-18} \\
    \tilde{d}_{x_i}'(\bm{k})&=\tilde{d}_{x_i}(\bm{k})\tilde{d}_{x_i}(\bm{k+b}). \label{eq:QD-19}
\end{align}
Meanwhile, one can show (see Appendix~\ref{sec:QA-1})
\begin{align}
    \tilde{d}_{x_i}(\Theta_{x_i}\bm{k})=\tilde{d}_{x_i}(\bm{k}), \label{eq:QD-20}
\end{align}
on the plane $k_{x_{i+1}}=\pi$. Then, substituting Eqs.~(\ref{eq:QD-17})-(\ref{eq:QD-20}) into Eq.~(\ref{eq:Q4-1}), we finally obtain Eq.~(\ref{eq:QD-14}).

\subsection{\label{sec:QD-2}Simplification of topological charge formulas}
We next show that additional symmetries relating two HSPs $W$ and $WA$ simplify the formulas of topological charges. As discussed in Sec.~\ref{sec:Q5-3}, there are some additional symmetries relating $\mathcal{GT}$-invariant HSPs $W$ and $WA$ in systems on a body-centered lattice. Then, such symmetries also relate $\mathcal{Q}_{xy}[S^2(W)]$ and $\mathcal{Q}_{xy}[S^2(WA)]$. Moreover, as a consequence, they simplify the form of $\mathcal{Q}_{xy}[T]=\mathcal{Q}_{xy}[T']$.

Firstly, under such symmetries, the $Z_2\times Z_2$ charges are related as
\begin{align}
    \mathcal{Q}_{xy}^{(i)}[S^2(WA)]=\mathcal{Q}_{xy}^{(i)}[S^2(W)]+
    \begin{cases}
    \mathcal{Q}[S^2(W)]\ \ \ & \text{for  $A$}, \\
    0\ \ \ & \text{for $B$,} 
    \end{cases} \label{eq:QD-2-0}
\end{align}
where groups $A/B$ are the classification of MSGs shown in Sec.~\ref{sec:Q5-3}. Therefore, symmetries in the group $A$ force the value of the $Z_2\times Z_2$ charge $\mathcal{Q}_{xy}[S^2]$ for the Dirac point at $W$ and that at $WA$ to be different, and symmetries in the group $B$ force them to be the same.
To show Eq.~(\ref{eq:QD-2-0}), we here focus on MSG \#45.236, and set $G=C_{2z}\mathcal{T}$.  Firstly, from the relation $\tilde{\Theta}_xG=(-1)^f\{E|110\}G\tilde{\Theta}_x$, we have
\begin{align}
    \frac{\mathrm{Pf}[\omega_x(K_0')]\mathrm{Pf}[\omega_x(K_0)]}{\mathrm{det}[\omega_G(K_0)]}&=(-1)^{\frac{N_{\mathrm{occ}}}{2}f}e^{-i\frac{N_{\mathrm{occ}}}{2}r}, \label{eq:QD-2-1} \\
    \frac{\tilde{d}_x(G\bm{k})\tilde{d}_x(\bm{k})}{\tilde{d}_G(\tilde{\Theta}_x\bm{k})\tilde{d}_G(\bm{k})}&=e^{-i\frac{N_{\mathrm{occ}}}{2}(k_x-k_y)}. \label{eq:QD-2-2}
\end{align}
Similar discussions hold for $\tilde{\Theta}_y$. Then, using Eqs.~(\ref{eq:QD-2-1}) and (\ref{eq:QD-2-2}) and corresponding equations for $\tilde{\Theta}_y$, one can see
\begin{align}
    &(-1)^{\mathcal{Q}_{xy}^{(0)}[S^2(WA)]} \notag \\
    &=i^{n_-(Q_{WA})+n_-(Q_W)}(-1)^{\mathcal{Q}_{xy}^{(0)}[S^2(W)]+\frac{N_{\mathrm{occ}}}{2}f}. \label{eq:QD-2-3}
\end{align}
Here, when $f=0$, since $G$ do not exchange the sigh of the $C_{2z}$-eigenvalue of occupied bands, we have $n_-(Q_{WA})+n_-(Q_W)=2n_-(Q_W)$. Meanwhile, when $f=1$, since $G$ exchange the sigh of the $C_{2z}$-eigenvalue of occupied bands, we have $n_-(Q_{WA})+n_-(Q_W)=n_+(Q_W)+n_-(Q_W)=N_{\mathrm{occ}}$. Moreover, we also have $n_-(Q_W)=\mathcal{Q}[S^2(W)]$ from Eq.~(\ref{eq:Q3-7}). Then, combining these equations, we finally obtain Eq.~(\ref{eq:QD-2-0}). Almost the same discussions hold for other MSGs by setting $G=C_{2z}\{E|\frac{1}{2}\frac{1}{2}0\}'$ for \#44.233, $G=C_{2z}\{E|0\frac{1}{2}0\}'$ for \#46.248, $G=M_z$ for \#72.543, $G=\{M_z|0\frac{1}{2}0\}$ for \#73.551, $G=C_{4z}$ for \#108.237, $G=\{C_{4z}|0\frac{1}{2}\frac{1}{4}\}$ for \#110.249, and $G=\{\bar{C_{4z}}|0\}'$ for \#120.323.

Here, we can show symmetries in the group $A(B)$ force the value of $\mathcal{Q}_{xy}[S^2]$ for the Dirac point at $W$ and that at $WA$ to be different (same) in a different way. MSGs in the groups $A/B$ only have 2D irreps at $W$ and $WA$. Therefore, there exists a symmetry-preserving perturbation that splits the Dirac point at $W$ into two Weyl points at $(\pi,\pi,\pi+\delta)$ and at $(\pi,\pi,\pi-\delta)$, and the Dirac point at $WA$ into two Weyl points at $(\pi,\pi,-\pi+\delta)$ and at $(\pi,\pi,-\pi-\delta)$, with a constant $\delta>0$. Then, after adding the perturbation, MSGs in $A$ force the monopole charge $C(=\pm 1)$ for the Weyl point at $(\pi,\pi,\pi+\delta)$ and that at  $(\pi,\pi,-\pi+\delta)$ to be the same but $\Delta n_-$ to be different, or $C$ to be different but $\Delta n_-$ to be the same. Therefore, as discussed in Sec.~\ref{sec:Q4-2}, Dirac points at $W$ and $WA$ have different $\mathcal{Q}_{xy}[S^2]$ in this case. Meanwhile, MSGs in $B$ force $C$ to be the same and $\Delta n_-$ to be the same, or $C$ to be different and $\Delta n_-$ to be different. Therefore, Dirac points at $W$ and $WA$ have the same $\mathcal{Q}_{xy}[S^2]$ in this case.

Next, as a consequence of the result, the form of $\mathcal{Q}_{xy}[T]$ is simplified. In fact,  by using Eqs.~(\ref{eq:QD-0}), (\ref{eq:QD-6}) and (\ref{eq:QD-14}), $\mathcal{Q}_{xy}[T]$ can be expressed without using $\mathcal{Q}_{xy}[S^2]$ as
\begin{align}
    \mathcal{Q}_{xy}[T]=
    \begin{cases}
        \mathcal{Q}[S^2(W)]+\frac{C[S^2(X)]}{2} & \text{for $A_+$}, \\
        \frac{N_{\mathrm{occ}}}{2} & \text{for $A_-$}, \\
        0 & \text{for $B_+$}, \\
        \mathcal{Q}[S^2(W)]+\frac{C[S^2(X)]+N_{\mathrm{occ}}}{2} & \text{for $B_-$}.
    \end{cases}
    \label{eq:QD-2-4}
\end{align}
Here, $(A/B)_{+}$ are subgroups of the groups $A/B$ which do not exchange the sign of the $C_{2z}$-eigenvalue of occupied bands, and $(A/B)_{-}$ are subgroups of the groups $A/B$ which exchange it. Explicitly, $A_+$ contains spinless \#44.233, spinless \#45.236, spinful \#46.248, $A_-$ contains \#73.551 and \#110.249, $B_+$ contains  \#72.543 and \#108.237, and $B_-$ contains spinful \#44.233, spinful \#45.236, spinless \#46.248, and spinful \#120.323.

We finally note that, when there are only two gapless nodes at $W$ and $WA$, we obtain Eqs.~(\ref{eq:Q5-3-1}) and (\ref{eq:Q5-3-3}) from Eq.~(\ref{eq:QD-2-4}). In fact, in this case, we have $\mathcal{Q}[S^2(W)]=n_-(X)$ from Eq.~(\ref{eq:Q3-7}), and $C[S^2(X)]=0$. Moreover, by considering the irreducible co-representations at $X$, one can see $n_-(X)=\frac{N_{\mathrm{occ}}}{2}\ (\mathrm{mod}\ 2)$ for $(A/B)_-$.

\section{\label{sec:QE}Some notes on the tight-binding model in Sec.~\ref{sec:Q6}}
Firstly, we give an explicit form of the tight-binding model $H_{\mathrm{TBM}}(\bm{k})$ and $\Delta(\bm{k})$ discussed in Sec.~\ref{sec:Q6}. We put four sites $A_1: (0,0,0)$, $A_2: (\frac{1}{2}, \frac{1}{2}, 0)$, $A_3: (\frac{1}{2},0,0)$, and $A_4: (0,\frac{1}{2},0)$ in the body-centered orthorhombic lattice, and take the basis set $[\psi_{A_1}, \psi_{A_2}, \psi_{A_3}, \psi_{A_4}]=[\psi_{A_1}, \tilde{\Theta}_x\psi_{A_1}, \tilde{\Theta}_y\psi_{A_1}, \tilde{\Theta}_x\tilde{\Theta}_y\psi_{A_1}]$, with $\mathcal{P}\psi_{A_1}=\psi_{A_1}$. Then, by introducing real hopping between the sites, we obtain a 3D tight-binding model $H_{\mathrm{TBM}}(\bm{k})$ and the perturbation term $\Delta(\bm{k})$ given by 
\begin{widetext}
 \begin{align}
    H_{\mathrm{TBM}}(\bm{k})&=\mqty(\sin(\frac{k_x+k_y}{2})\sin(\frac{k_z}{2}) & ie^{-i\frac{k_x+k_y}{2}}\cos(\frac{k_z}{2}) & s(1+e^{-ik_x}) & t(1+e^{-ik_y}) \\ -ie^{i\frac{k_x+k_y}{2}}\cos(\frac{k_z}{2}) & -\sin(\frac{k_x-k_y}{2})\sin(\frac{k_z}{2}) & t(1+e^{ik_y}) & s(1+e^{ik_x}) \\ s(1+e^{ik_x}) & t(1+e^{-ik_y}) & \sin(\frac{k_x-k_y}{2})\sin(\frac{k_z}{2}) & -ie^{i\frac{k_x-k_y}{2}}\cos(\frac{k_z}{2}) \\ t(1+e^{ik_y}) & s(1+e^{-ik_x}) & ie^{-i\frac{k_x-k_y}{2}}\cos(\frac{k_z}{2}) & -\sin(\frac{k_x+k_y}{2})\sin(\frac{k_z}{2})), \\
    \Delta(\bm{k})&=iu\sin(\frac{k_x}{2})\sin(\frac{k_z}{2})\mqty(0 & 0 & 0 & e^{-i\frac{k_y}{2}} \\ 0 & 0 & e^{i\frac{k_y}{2}} & 0 \\ 0 & -e^{-i\frac{k_y}{2}} & 0 & 0 \\ -e^{i\frac{k_y}{2}} & 0 & 0 & 0),
\end{align}
\end{widetext}
under the basis set. Here $s,t$, and $u$ are real parameters, and we set $s=0.3, t=0.3$, and $u=0.2$ in the main text.

Next, in this model, $\tilde{\Theta}_x,\ \tilde{\Theta}_y,$ and $ \mathcal{P}$ symmetries are represented by
\begin{align}
\tilde{\Theta}_x(\bm{k})&=\mqty(0 & e^{ik_y} & 0 & 0 \\ 1 & 0 & 0 & 0 \\ 0 & 0 & 0 & e^{ik_y} \\ 0 & 0 & 1 & 0)K, \\
\tilde{\Theta}_y(\bm{k})&=\mqty(0 & 0& e^{ik_x} & 0 \\ 0 & 0 &  0 & e^{ik_y} \\ 1 & 0 & 0 & 0 \\ 0 & e^{ik_x+ik_y} & 0 & 0)K, \\
\mathcal{P}(\bm{k})&=\mqty( 1 & 0 & 0 & 0 \\ 0 & e^{-ik_x-ik_y} & 0 & 0 \\ 0 & 0 & e^{-ik_x} & 0 \\ 0 & 0 & 0 & e^{-ik_y}),
\end{align}
respectively, where $K$ is the complex conjugate operator.

\begin{figure}
\includegraphics[width=\columnwidth, pagebox=cropbox, clip]{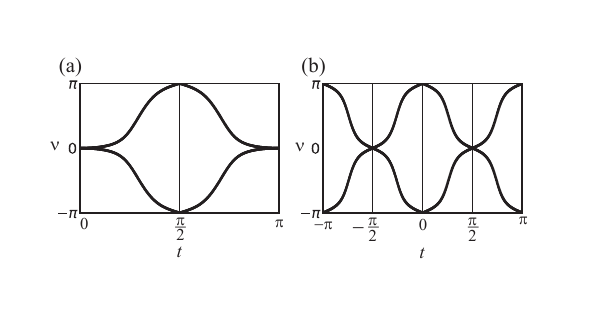}
\caption{\label{fig:QD}\small{Wilson loop spectra. (a) Wilson loop spectrum for $\mathcal{Q}[S^2]$. The bands intersect a reference line $\nu=\mathrm{const.}$ odd times within $0\le t \le \frac{\pi}{2}$. (b) Wilson loop spectrum for $\mathcal{Q}_{xy}[T']$. The bands intersect a reference line $\nu=\mathrm{const.}$ odd times within $0\le t \le \frac{\pi}{2}$.}}
\end{figure}

Lastly, we calculate the value of $Z_2$ charges for the system. Firstly, the value of $\mathcal{Q}[S^2]$ with $S^2=\{\bm{k}| k_x=\pi+0.5\sin t\cos\phi, \ k_y=\pi+0.5\cos t,\ k_z=\pi+0.5\sin t\sin\phi\ \ (0\le\phi\le 2\pi, 0\le t\le \pi)\}$ is calculated by using the Wilson loop method discussed in Ref.~\cite{Yukitake2025}. The resulting spectrum is shown in Fig.~\ref{fig:QD}(a). Since the bands intersect a reference line $\nu=\mathrm{const.}$ odd times within $0\le t \le \frac{\pi}{2}$, we find $\mathcal{Q}[S^2]=1\ (\mathrm{mod}\ 2)$. Next, the value of $\mathcal{Q}_{xy}[T]=\mathcal{Q}_{xy}[T']$ (see Appendix~\ref{sec:QD-1}) with $T'=\{\bm{k} | k_x=\pi+0.5\cos t,\ k_y=\pi+0.5\sin t,\ k_z=k  \ \ (-2\pi\le k\le 2\pi\ ,-\pi\le t\le\pi)\}$ is calculated by using the Wilson loop method for $Z_2$ topological insulators discussed in Ref.~\cite{Yu2011}. The resulting spectrum is shown in Fig.~\ref{fig:QD}(b). Here, in our case, one can easily see from the discussion in Ref.~\cite{Yu2011} that $\mathcal{Q}_{xy}[T']=1\ (\mathrm{mod}\ 2)$ if and only if the bands intersect a reference line $\nu=\mathrm{const.}$ odd times within $0\le t \le \frac{\pi}{2}$. Therefore, we find $\mathcal{Q}_{xy}[T']=1\ (\mathrm{mod}\ 2)$.


\begin{thebibliography}{0}%
\makeatletter
\providecommand \@ifxundefined [1]{%
 \@ifx{#1\undefined}
}%
\providecommand \@ifnum [1]{%
 \ifnum #1\expandafter \@firstoftwo
 \else \expandafter \@secondoftwo
 \fi
}%
\providecommand \@ifx [1]{%
 \ifx #1\expandafter \@firstoftwo
 \else \expandafter \@secondoftwo
 \fi
}%
\providecommand \natexlab [1]{#1}%
\providecommand \enquote  [1]{``#1''}%
\providecommand \bibnamefont  [1]{#1}%
\providecommand \bibfnamefont [1]{#1}%
\providecommand \citenamefont [1]{#1}%
\providecommand \href@noop [0]{\@secondoftwo}%
\providecommand \href [0]{\begingroup \@sanitize@url \@href}%
\providecommand \@href[1]{\@@startlink{#1}\@@href}%
\providecommand \@@href[1]{\endgroup#1\@@endlink}%
\providecommand \@sanitize@url [0]{\catcode `\\12\catcode `\$12\catcode `\&12\catcode `\#12\catcode `\^12\catcode `\_12\catcode `\%12\relax}%
\providecommand \@@startlink[1]{}%
\providecommand \@@endlink[0]{}%
\providecommand \url  [0]{\begingroup\@sanitize@url \@url }%
\providecommand \@url [1]{\endgroup\@href {#1}{\urlprefix }}%
\providecommand \urlprefix  [0]{URL }%
\providecommand \Eprint [0]{\href }%
\providecommand \doibase [0]{https://doi.org/}%
\providecommand \selectlanguage [0]{\@gobble}%
\providecommand \bibinfo  [0]{\@secondoftwo}%
\providecommand \bibfield  [0]{\@secondoftwo}%
\providecommand \translation [1]{[#1]}%
\providecommand \BibitemOpen [0]{}%
\providecommand \bibitemStop [0]{}%
\providecommand \bibitemNoStop [0]{.\EOS\space}%
\providecommand \EOS [0]{\spacefactor3000\relax}%
\providecommand \BibitemShut  [1]{\csname bibitem#1\endcsname}%
\let\auto@bib@innerbib\@empty
\end{thebibliography}%


\begin{thebibliography}{99}
\bibitem{Murakami2007}
S. Murakami, Phase transition between the quantum spin Hall and insulator phases in 3D: emergence of a topological gapless phase, New J. Phys. $\bm{9}$, 356 (2007).

\bibitem{Wan2011}
X. Wan, A. M. Turner, A. Vishwanath, and S. Y. Savrasov, Topological semimetal and Fermi-arc surface states in the electronic structure of pyrochlore iridates, Phys. Rev. B $\bm{83}$, 205101 (2011).

\bibitem{Fang2012}
C. Fang, M. J. Gilbert, X. Dai, and B. A. Bernevig, Multi-Weyl Topological Semimetals Stabilized by Point Group Symmetry, Phys. Rev. Lett. $\bm{108}$, 266802 (2012).

\bibitem{Li2015}
S. Li and A. V. Andreev, Spiraling Fermi arcs in Weyl materials, Phys. Rev. B $\bm{92}$, 201107(R) (2015).

\bibitem{Xu2015}
S.-Y. Xu, I. Belopolski, N. Alidoust, M. Neupane, G. Bian, C. Zhang, R. Sankar, G. Chang, Z. Yuan, C.-C. Lee, et al., Discovery of a Weyl fermion semimetal and topological Fermi arcs, Science $\bm{349}$, 613 (2015).

\bibitem{Lv2015}
B. Q. Lv, H. M. Weng, B. B. Fu, X. P. Wang, H. Miao, J. Ma, P. Richard, X. C. Huang, L. X. Zhao, G. F. Chen, Z. Fang, X. Dai, T. Qian, and H. Ding, Experimental Discovery of Weyl Semimetal TaAs, Phys. Rev. X $\bm{5}$, 031013 (2015).

\bibitem{Tsirkin2017}
S. S. Tsirkin, I. Souza, and D. Vanderbilt, Composite Weyl nodes stabilized by screw symmetry with and without time-reversal invariance, Phys. Rev. B $\bm{96}$, 045102 (2017).

\bibitem{HWang2020}
H.-X. Wang, Z.-K. Lin, B. Jiang, G.-Y. Guo, and J.-H. Jiang, Higher-Order Weyl Semimetals, Phys. Rev. Lett. $\bm{125}$, 146401 (2020).

\bibitem{Young2012}
S. M. Young, S. Zaheer, J. C. Y. Teo, C. L. Kane, E. J. Mele, and A. M. Rappe, Dirac Semimetal in Three Dimensions, Phys. Rev. Lett. $\bm{108}$, 140405 (2012).

\bibitem{Yang2014}
B.-J. Yang, and N. Nagaosa, Classification of stable three-dimensional Dirac semimetals with nontrivial topology, Nat. Commun. $\bm{5}$, 4898 (2014).

\bibitem{Yang2015}
B.-J. Yang, T. Morimoto, and A. Furusaki, Topological charges of three-dimensional Dirac semimetals with rotation symmetry, Phys. Rev. B $\bm{92}$, 165120 (2015).

\bibitem{Gao2016}
Z. Gao, M. Hua, H. Zhang, and X. Zhang, Classification of stable Dirac and Weyl semimetals with reflection and rotational symmetry, Phys. Rev. B $\bm{93}$, 205109 (2016).

\bibitem{Wang2012}
Z. Wang, Y. Sun, X.-Q. Chen, C. Franchini, G. Xu, H. Weng, X. Dai, and Z. Fang, Dirac semimetal and topological phase transitions in A$_3$Bi (A=Na, K, Rb), Phys. Rev. B $\bm{85}$, 195320 (2012).

\bibitem{Liu2014}
Z. K. Liu, B. Zhou, Y. Zhang, Z. J. Wang, H. M. Weng, D. Prabhakaran,  S.-K. Mo, Z. X. Shen, Z. Fang, X. Dai, Z. Hussain, Y. L. Chen, Discovery of a Three-Dimensional Topological Dirac Semimetal, Na$_3$Bi, Science $\bm{343}$, 864 (2014).

\bibitem{Xu2014}
S.-Y. Xu, C. Liu, S. K. Kushwaha, R. Sankar, J. W. Krizan, I. Belopolski, M. Neupane, G. Bian, N. Alidoust, T.-R. Chang, et al., Observation of Fermi arc surface states in a topological metal, Science $\bm{347}$, 6219 (2014).

\bibitem{Wang2013}
Z. Wang, H. Weng, Q. Wu, X. Dai, and Z. Fang, Three-dimensional Dirac semimetal and quantum transport in Cd$_3$As$_2$, Phys. Rev. B $\bm{88}$, 125427 (2013).

\bibitem{Borisenko2014}
S. Borisenko, Q. Gibson, D. Evtushinsky, V. Zabolotnyy,
B. B$\ddot{\mathrm{u}}$chner, and R. J. Cava, Experimental Realization of a Three-Dimensional Dirac Semimetal, Phys. Rev. Lett. $\bm{113}$, 027603 (2014).

\bibitem{Morimoto2014}
T. Morimoto and A. Furusaki, Weyl and Dirac semimetals with $\mathbb{Z}_2$ topological charge, Phys. Rev. B $\bm{89}$, 235127 (2014).

\bibitem{Tang2016}
P. Tang, Q. Zhou, G. Xu, and S.-C. Zhang, Dirac fermions in an antiferromagnetic semimetal, Nat. Phys. $\bm{12}$, 1100 (2016).

\bibitem{Kargarian2016}
M. Kargarian, M. Randeria, and Y.-M. Lu, Are the surface Fermi arcs in Dirac semimetals topologically protected?, Proc. Natl. Acad. Sci. USA $\bm{113}$, 8648 (2016).

\bibitem{Le2018}
C. Le, X. Wu, S. Qin, Y. Li, R. Thomale, F.-C. Zhang, and J. Hu, Dirac semimetal in $\beta$-CuI without surface Fermi arcs, Proc. Natl. Acad. Sci. USA $\bm{115}$, 8311 (2018).

\bibitem{Kargarian2018}
M. Kargarian, Y.-M. Lu, and M. Randeria, Deformation and stability of surface states in Dirac semimetals, Phys. Rev. B $\bm{97}$, 165129 (2018).

\bibitem{Wu2019}
Y. Wu, N. H. Jo, L.-L. Wang, C. A. Schmidt, K. M. Neilson, B. Schrunk,
P. Swatek, A. Eaton, S. L. Bud’ko, P. C. Canfield, and Adam Kaminski, Fragility of Fermi arcs in Dirac semimetals, Phys. Rev. B $\bm{99}$, 161113(R) (2019).

\bibitem{Lin2020}
Z. Lin, C. Wang, P. Wang, S. Yi, L. Li, Q. Zhang, Y. Wang, Z. Wang, H. Huang, Y. Sun, et al., Dirac fermions in antiferromagnetic FeSn kagome lattices with combined space inversion and time-reversal symmetry, Phys. Rev. B $\bm{102}$, 155103 (2020).

\bibitem{Wieder2020}
B. J. Wieder, Z. Wang, J. Cano, X. Dai, L. M. Schoop, B. Bradlyn, and B. A. Bernevig, Strong and fragile topological Dirac semimetals with higher-order Fermi arcs, Nat. Commun. $\bm{11}$, 627 (2020).

\bibitem{Fang2021}
Y. Fang and J. Cano, Classification of Dirac points with higher-order Fermi arcs, Phys. Rev. B $\bm{104}$, 245101 (2021).

\bibitem{Xia2022}
C.-H. Xia, H.-S. Lai, X.-C. Sun, C. He, and Y.-F. Chen, Experimental Demonstration of Bulk-Hinge Correspondence in a Three-Dimensional Topological Dirac Acoustic Crystal, Phys. Rev. Lett. $\bm{128}$, 115701 (2022).

\bibitem{Qian2023}
S. Qian, Y. Li, and C.-C. Liu, Stable higher-order topological Dirac semimetals with $\mathbb{Z}_2$ monopole charge in alternating-twist multilayer graphene and beyond, Phys. Rev. B $\bm{108}$, L241406 (2023).

\bibitem{Burkov2011}
A. A. Burkov, M. D. Hook, and L. Balents, Topological nodal semimetals, Phys. Rev. B $\bm{84}$, 235126 (2011).

\bibitem{Fang2015}
C. Fang, Y. Chen, H.-Y. Kee, and L. Fu, Topological nodal line semimetals with and without spin-orbital coupling, Phys. Rev. B $\bm{92}$, 081201(R) (2015).

\bibitem{Chen2016}
Y. Chen, H.-S. Kim, and H.-Y. Kee, Topological crystalline semimetals in nonsymmorphic lattices, Phys. Rev. B $\bm{93}$, 155140 (2016).

\bibitem{Zhao2017}
Y. X. Zhao, and Y. Lu, $PT$-Symmetric Real Dirac Fermions and Semimetals, Phys. Rev. Lett. $\bm{118}$, 056401 (2017).

\bibitem{Li2018}
S. Li, Y. Liu, S.-S. Wang, Z.-M. Yu, S. Guan, X.-L. Sheng, Y. Yao, and S. A. Yang, Nonsymmorphic-symmetry-protected hourglass Dirac loop, nodal line, and Dirac point in bulk and
monolayer $X_3$SiTe$_6$ ($X=$Ta, Nb), Phys. Rev. B $\bm{97}$, 045131 (2018).

\bibitem{Ahn2018}
J. Ahn, D. Kim, Y. Kim, and B.-J. Yang, Band Topology and Linking Structure of Nodal Line Semimetals with $Z_2$ Monopole Charges, Phys. Rev. Lett. $\bm{121}$, 106403 (2018).

\bibitem{Bouhon2019}
A. Bouhon, A. M. Black-Schaffer, and R.-J. Slager, Wilson loop approach to fragile topology of split elementary band representations and topological crystalline insulators with time-reversal symmetry, Phys. Rev. B $\bm{100}$, 195135 (2019).

\bibitem{Sheng2019}
X.-L. Sheng, C. Chen, H. Liu, Z. Chen, Z.-M. Yu, Y. X. Zhao, and S. A. Yang, Two-Dimensional Second-Order Topological Insulator in Graphdiyne, Phys. Rev. Lett. $\bm{123}$, 256402 (2019).

\bibitem{Wang2020}
K. Wang, J.-X. Dai, L.B. Shao, S. A. Yang, and Y. X. Zhao, Boundary Criticality of $PT$-Invariant Topology and Second-Order Nodal-Line Semimetals, Phys. Rev. Lett. $\bm{125}$, 126403 (2020).

\bibitem{Lee2020}
E. Lee, R. Kim, J. Ahn, and B.-J. Yang, Two-dimensional higher-order topology in monolayer graphdiyne, npj Quantum Mater. $\bm{5}$, 1 (2020).

\bibitem{Chen2021}
C. Chen, W. Wu, Z.-M. Yu, Z. Chen, Y. X. Zhao,  X.-L. Sheng,
and S. A. Yang, Graphyne as a second-order and real Chern topological insulator in two dimensions, Phys. Rev. B $\bm{104}$, 085205 (2021).

\bibitem{Chen2022}
C. Chen, X.-T. Zeng, Z. Chen, Y. X. Zhao, X.-L. Sheng, and S. A. Yang, Second-Order Real Nodal-Line Semimetal in Three-Dimensional Graphdiyne, Phys. Rev. Lett. $\bm{128}$, 026405 (2022).

\bibitem{Xue2023}
H. Xue, Z. Y. Chen, Z. Cheng, J. X. Dai, Y. Long, X. Y. Zhao, and B. Zhang, Stiefel-Whitney topological charges in a three-dimensional acoustic nodal-line crystal, Nat. Commun. $\bm{14}$, 4563 (2023).

\bibitem{Xiang2024}
X. Xiang, Y.-G. Peng, F. Gao, X. Wu, P. Wu,
Z. Chen, X. Ni, and X.-F. Zhu, Demonstration of Acoustic Higher-Order Topological Stiefel-Whitney Semimetal, Phys. Rev. Lett. $\bm{132}$, 197202 (2024).

\bibitem{Ma2024}
Q. Ma, Z. Pu, L. Ye, J. Lu, X. Huang, M. Ke, H. He,  W. Deng, and Z. Liu, Observation of Higher-Order Nodal-Line Semimetal in Phononic Crystals, Phys. Rev. Lett. $\bm{132}$, 066601 (2024).

\bibitem{Wang2024}
X. Wang, J. Bai, J. Wang, Z. Cheng, S. Qian, W. Wang, G. Zhang, Z.-M. Yu, and Y. Yao, Real Topological Phonons in 3D Carbon Allotropes, Adv. Mater. $\bm{36}$, 2407437 (2024).

\bibitem{Yue2024}
S. J. Yue, Q. Liu, S. A. Yang, and Y. X. Zhao, Stability and noncentered $PT$ symmetry of real topological phases, Phys. Rev. B $\bm{109}$, 195116 (2024).

\bibitem{Fang2016}
C. Fang, L. Lu, J. Liu, and L. Fu, Topological semimetals with helicoid surface states, Nat. Phys. $\bm{12}$, 936 (2016).

\bibitem{Cheng2020}
H. Cheng, Y. Sha, R. Liu, C. Fang, and L. Lu, Discovering Topological Surface States of Dirac Points, Phys. Rev. Lett. $\bm{124}$, 104301 (2020).

\bibitem{Cai2020}
X. Cai, L. Ye, C. Qiu, M. Xiao, R. Yu, M. Ke, and Z. Liu, Symmetry-enforced three-dimensional Dirac phononic crystals, Light Sci. Appl. $\bm{9}$, 38 (2020).

\bibitem{Su2022}
Z. Su, W. Gao, B. Liu, L. Huang, and Y. Wang, Three-dimensional Dirac semimetal metamaterial enabled by negative couplings, New J. Phys. $\bm{24}$, 033025 (2022).

\bibitem{Zhang2022}
T. Zhang, D. Hara, and S. Murakami, $Z_2$ Dirac points with topologically protected multihelicoid surface states, Phys. Rev. Res. $\bm{4}$, 033170 (2022).

\bibitem{Hara2023}
D. Hara, Interplay between topological surface states in topological semimetals, Ph.D. thesis, Tokyo Institute of Technology, 2023.

\bibitem{Zhang2023}
T. Zhang and S. Murakami, Parallel and anti-parallel helical surface states for topological semimetals, Sci. Rep. $\bm{13}$, 9239 (2023).

\bibitem{Yukitake2025}
T. Yukitake, D. Hara, S. Murakami, Double helicoid surface states in Dirac semimetals protected by glide-time-reversal symmetry, Phys. Rev. B $\bm{112}$, 045137 (2025).

\bibitem{FuKane2006}
L. Fu and C. L. Kane, Time reversal polarization and a $Z_2$ adiabatic spin pump, Phys. Rev. B $\bm{74}$, 195312 (2006).

\bibitem{FuKane2007}
L. Fu and C. L. Kane, Topological insulators with inversion symmetry, Phys. Rev. B $\bm{76}$, 045302 (2007).

\bibitem{Graf2013}
G. Graf and M. Porta, Bulk-edge correspondence for two-dimensional topological insulators, Commun. Math. Phys. $\bm{324}$, 851 (2013).

\bibitem{Hatcher2001}
A. Hatcher, \textit{Algebraic topology} (Cambridge University Press, Cambridge 2002).

\bibitem{Kim2020}
H. Kim and S. Murakami, Glide-symmetric topological crystalline insulator phase in a nonprimitive lattice, Phys. Rev. B $\bm{102}$, 195202 (2020).

\bibitem{Yu2011}
R. Yu, X. L. Qi, A. Bernevig, Z. Fang, and X. Dai, Equivalent expression of $\mathbb{Z}_2$ topological invariant for band insulators using the non-Abelian Berry connection, Phys. Rev. B $\bm{84}$, 075119 (2011).

\end{thebibliography}
\end{document}